\documentclass{article}
\usepackage[small]{diagrams}

\usepackage{amsfonts}

\newcommand{\bb}{\begin{eqnarray}}
\newcommand{\ee}{\end{eqnarray}}
\newcommand{\beq}{\begin{equation}}
\newcommand{\eeq}{\end{equation}}
\newcommand{\ba}{\begin{array}}
\newcommand{\ea}{\end{array}}
\newcommand{\bd}{\begin{displaymath}}
\newcommand{\ed}{\end{displaymath}}

\def\br{{[}}

\def\bbbc{{\mathbb C}}

\def\bbbz{{\mathbb Z}}

\def\bbbd{{\mathbb D}}
\def\bbbt{{\mathbb T}}
\def\bbbo{{\mathbb O}}
\def\bbbi{{\mathbb I}}
\def\cp1{{\mathbb C\mathbb P}^1}
\def\cA{{\cal A}}

\def\cL{{\cal L}}

\def\cR{{\cal R}}
\def\cG{{\cal G}}

\def\l{{\lambda}}
\def\ord{{\mbox{ord}}}
\def\aut{{\mbox{Aut\,}}}

\def\e{{\bf e}}
\newtheorem{theorem}{Theorem}[section]
\newtheorem{lemma}[theorem]{Lemma}

\newtheorem{prop}[theorem]{Proposition}

\newtheorem{df}[theorem]{Definition}

\newcommand{\pf}{\textbf{Proof }}

\title{Reduction Groups and Automorphic Lie Algebras.}
\author{S. Lombardo$^{1}$, A.V. Mikhailov$^{2}$\footnote{L.D.Landau Institute for Theoretical Physics Chernogolovka, Russia (on leave)}\\
 Department of Applied Mathematics\\
University of Leeds, Leeds LS2 9JT, UK\\
\small E-mail addresses: $^{1}$ sara@maths.leeds.ac.uk,
$^{2}$ sashamik@maths.leeds.ac.uk}

\date{}

\begin{document}
\maketitle

\begin{abstract}
We study a new class of infinite dimensional
Lie algebras, which has important applications to the theory of integrable equations. 
The construction of these algebras is very similar to the
one for automorphic functions and this motivates the name {\em automorphic
Lie algebras}. For automorphic Lie algebras we present bases in which
they are quasigraded and all structure constants can be written out
explicitly. These algebras have a useful factorisations on two subalgebras
similar to the factorisation of the current algebra on the positive and
negative parts.
\end{abstract}

\section{Introduction}

In this paper we introduce and study automorphic Lie algebras.  This
subclass of infinite dimensional Lie algebras is very useful for
applications and actually has been motivated by applications to the
theory of integrable equations.  Automorphic Lie algebras are
quasigraded and all their structure constants can be found
explicitly. They form a more general class than graded infinite dimensional
Lie algebras \cite{kac}, they also have rich internal structure and can
be studied in depth.

The basic construction is very similar to the theory of automorphic
functions \cite{klein}, \cite{ford}. In a sense, it is a generalisation
of this theory to the case of semi--simple Lie algebras over a ring of
meromorphic functions $\cR (\Gamma)$ of a complex parameter $\lambda$
with poles in a set of points $\Gamma$. Suppose $\cG$ is a discontinuous group of
fractional-linear transformations of the complex variable $\lambda$ and the set 
$\Gamma$ is an orbit of this group or a finite union of orbits, then
transformations from $\cG$ induce automorphisms of the ring $\cR
(\Gamma)$. A set of elements of $\cR (\Gamma)$ which are invariant
with respect to $\cG$ form a subring of automorphic functions.
Automorphic algebras are defined in a very similar way. Let us consider a finite dimensional semi-simple Lie algebra $\cA $ over the
ring  $\cR (\Gamma)$. This algebra can be viewed as an infinite dimensional Lie algebra over $\bbbc$ and will be
denoted  $\cA (\Gamma)$.  Suppose $G$ is a
subgroup of the group of automorphisms of $\cA(\Gamma)$. Elements of $G$ are simultaneous
transformations (automorphisms) of the semi-simple Lie algebra $\cA$ and the ring
$\cR (\Gamma)$. Then the automorphic Lie algebra $\cA_G (\Gamma)$
is defined as the set of all
elements of $\cA (\Gamma)$ which are $G$ invariant.

In this paper we restrict ourselves to finite groups of fractional-linear transformations of the Riemann sphere and therefore the set
$\Gamma$ is finite and all elements of $\cR (\Gamma)$ are rational
functions. The theory of automorphic
functions  for finite groups has been developed by Felix Klein \cite{klein}, \cite{klein1}; automorphic functions corresponding to finite groups can be easily obtained using the group average. The paper is organised as follows: 
in the second section we introduce notations and recall
some useful results from the theory of 
elementary automorphic functions. We give a brief account of automorphisms
of semi-simple Lie algebras, discuss the structure of automorphisms
groups of algebras over a ring of rational functions and define
automorphic Lie algebras. In the third section we construct explicitly 
 automorphic Lie algebras corresponding to the dihedral
group $\bbbd_{N}$ and study some of their properties.  In particular we  built
explicitly bases in which these algebras are quasigraded and
find all structure constants. The group of automorphisms of a semi-simple Lie algebra is a continuous Lie group and therefore its elements may depend on the complex parameter $\lambda$. In this case the reduction group $G$ is a subgroup of a semi-direct product of $\cG$ and $\aut{\cA}$. A nontrivial example of the corresponding automorphic Lie algebra is given in section \ref{semidirect}. For completeness, in the Appendix we give an account of all finite groups of fractional linear transformations,
their orbits and primitive automorphic functions.

Originally our study has been motivated by the problem of reduction of
Lax pairs.  Most of integrable
equations interesting for applications are results of reductions of bigger systems. The problem of
reductions is one of the central problems in the theory of integrable
equations. A wide class of algebraic reductions can be studied in terms
of {\em reduction groups}. The concept of reduction group has been
formulated in \cite{AM0}, \cite{AM1}, \cite{AM2} and developed in
\cite{AM4}, \cite{ggk}, \cite{G}, \cite{msy1}. It has been successfully
applied and proved to be very useful for a classification  of solutions
of the classical Yang-Baxter equation \cite{bel}, \cite{drbel}. The
most recent publications related to the reduction group are \cite{lm1}, \cite{sokgol}.

A reduction group $G$ is a discrete group of automorphisms of a Lax pair. Its
elements are simultaneous gauge transformations and fractional-linear
transformations of the spectral parameter. The requirement that a Lax
pair is invariant with respect to a reduction group imposes certain
constraints on the entries of the Lax pair and yields a reduction.
Simultaneous gauge transformations and fractional-linear
transformations of the spectral parameter are automorphisms of 
the underlying infinite dimensional Lie algebra $\cA(\Gamma)$.
The reduction corresponding to $G$ is nothing but a restriction of the Lax pair to the automorphic subalgebra $\cA_G(\Gamma)\subset \cA(\Gamma)$.

About a year ago we discussed our new developments in the theory of
reductions and reduction groups \cite{lm1} with V.V. Sokolov, who
suggested us to reformulate our results in algebraic terms in order to
make them accessible to a wider mathematical community.  We are
grateful to him for this advise. Indeed, Lie algebras have applications
far beyond the theory of integrable equations. We believe
 \emph{automorphic Lie algebras} are a new and
important class of infinite dimensional Lie algebras which deserves
further study and development.

\section{Automorphisms}

\subsection{Finite groups of automorphisms of the complex plane and rational automorphic functions}

Let $\hat{\cG}$ be a group of fractional-linear transformations $\sigma _r$
\beq\label{fltr}
  \lambda_r=\sigma_r(\lambda)=\frac{a_r \lambda+b_r}{c_r \lambda+d_r}\, ,\quad a_r d_r-b_r c_r=1\, , \eeq
where $\sigma_0$ is the identity transformation ($id$) of the group
\[ \sigma_0(\lambda)=\lambda\, ,\quad a_0=d_0=\pm 1\, ,\quad b_0=c_0=0 \, .\]
The composition $\sigma_{r'}(\sigma_r(\lambda))$ defines the group product $\sigma_{r'}\cdot \sigma_{r}$.
We will denote $\sigma_r^{-1} (\lambda)$ the transformation inverse to $\sigma_r(\lambda)$.
One can associate $2\times 2$ matrices with fractional-linear transformations (\ref{fltr})
\[  \left(\ba{cc} a_r&b_r\\c_r&d_r\ea\right)\to \sigma_r \, .\]
The product of such matrices corresponds to the composition of fractional-linear transformations.
It defines a homomorphism of the group $SL(2,\bbbc)$ onto the group $\hat{\cG}$. The kernel of the homomorphism consists of two elements $I_2$ and $-I_2$ where $I_2$ is the unit $2\times 2$
matrix. In other words, the group $\hat{\cG}$ is isomorphic to $PSL(2,\bbbc)=SL(2,\bbbc)/\{\pm I_2\}$.

Two groups $\cG$ and $\cG'$ of fractional-linear transformations are equivalent
if there is a fractional-linear transformation $\tau$ such that for any $\sigma\in \cG$  
\[ \sigma'=\tau ^{-1} \sigma\tau\in \cG' \] 
and any element of $\cG'$ can be obtained in this way.

Finite subgroups of $\hat{\cG}$ have been completely classified by Felix Klein \cite{klein1}.
The complete list of finite groups of fractional-linear transformations consists of five elements:
\begin{equation}\label{fgroups} 
\bbbz _N,\quad \bbbd_N,\quad \bbbt,\quad \bbbo,\quad \bbbi,
\end{equation}
i.e. the additive group of integers modulo $N$, the group of a dihedron with $N$ corners,  
the tetrahedral, octahedral and icosahedral groups, respectively. In this paper we consider only finite groups of 
fractional-linear transformations.

Let $\gamma_0$ be a complex number (a point on the Riemann sphere $\cp1$), and let $\cG$ be a finite group 
of fractional-linear transformations. The \emph{orbit} $\cG (\gamma_0)$ is defined as the set of all 
images $\cG (\gamma_0)=\{ \sigma_r(\gamma_0)\, |\, \sigma_r\in\cG\}$. If two orbits $\cG(\gamma_1)$ and $\cG(\gamma_2)$
have non-empty intersection, they coincide. The point $\gamma_0$ is called 
a {\em fixed point} of a transformation $\sigma_r$ if $\sigma_r(\gamma_0)=\gamma_0$. Transformations for which 
the point $\gamma_0$ is fixed form a subgroup $\cG_{\gamma_0}\subset\cG$, called {\em isotropy
subgroup} of $\gamma_0$. The \emph{order} of the fixed point is defined as the order of its isotropy 
subgroup $\ord (\gamma_0)=|\cG_{\gamma_0}|$. The point $\gamma_0$ and 
the corresponding orbit
$\cG(\gamma_0)$ are called \emph{generic}, if the isotropy subgroup $\cG_{\gamma_0}$ is trivial, 
i.e. it consists of the identity transformation only. The orbit $\cG(\gamma_0)$, and so $\gamma_0$, is called
\emph{degenerated}, if $|\cG_{\gamma_0}|>1$. It follows from Lagrange Theorem that the number
of points in the orbit $\cG(\gamma_0)$ is equal to $|\cG|/|\cG_{\gamma_0}|$.

Given a rational function $f(\lambda)$ of the complex variable $\lambda$, the action of the group $\cG$ is defined as
\begin{equation}\label{sigmaf} 
\sigma_r :f(\lambda)\to f(\sigma_r^{-1} (\lambda)) \, ,
\end{equation}
or simply $\sigma_r (f(\lambda))= f(\sigma_r^{-1} (\lambda))$. A non-constant function $f(\lambda)$ is 
called {\em automorphic function} of the group $\cG$ if $\sigma_r(f(\lambda))=f(\lambda)$ for all $\sigma_r\in\cG$.
Automorphic functions take the same value at all points of an orbit $\cG(\gamma_0)$.

The following important fact holds; it has been perfectly known to Felix
Klein, but it was not formulated as a separate statement in his book \cite{klein}.

\begin{theorem}\label{primfunth} Let $\cG$ be a finite group of fractional-linear transformations, and be
$\cG(\gamma_1), \cG(\gamma_2)$ any two different orbits, then:
\begin{enumerate}
\item There exists a {\em primitive} automorphic function $f(\lambda, \gamma_1, \gamma_2)$ with poles of multiplicity 
$|\cG_{\gamma_1}|$ at points $\cG(\gamma_1)$ and zeros of multiplicity 
$|\cG_{\gamma_2}|$ at points $\cG(\gamma_2)$ and with no other poles or zeros. Function 
$f(\lambda, \gamma_1, \gamma_2)$ is defined uniquely, up to a constant multiplier.
\item Any rational automorphic function of the group $\cG$ is a rational 
function of the \emph{primitive} $f(\lambda,\gamma_1,\gamma_2)$.
\end{enumerate}
\end{theorem}

If $f(\lambda, \gamma_1,\gamma_2)$ is a primitive automorphic function, then 
\bb f(\lambda, \gamma_2,\gamma_1)&=& \frac{c_1}{f(\lambda, \gamma_1,\gamma_2)}\, ,\label{gam21}\\ \label{gam13}
f(\lambda, \gamma_1 ,\gamma_3)&=&c_2(f(\lambda, \gamma_1,\gamma_2)-f(\gamma_3, \gamma_1,\gamma_2))\, 
,\quad \gamma_3\not\in\cG(\gamma_1)\, ,\\ \label{gam34}
f(\lambda, \gamma_3,\gamma_4)&=&c_3 \frac{f(\lambda, \gamma_1,\gamma_2)-f(\gamma_4, \gamma_1,\gamma_2)}
{f(\lambda, \gamma_1,\gamma_2)-f(\gamma_3, \gamma_1,\gamma_2)}\, ,\quad \gamma_3,\gamma_4\not\in\cG(\gamma_1)\, ,
\ee
where $c_1,c_2,c_3$ are nonzero complex constants. Thus, it is sufficient to find one primitive automorphic 
function $f= f(\lambda,\gamma_1,\gamma_2)$ and all other rational automorphic functions will be rational
functions of $f$.

For finite groups, automorphic functions can be obtained using the group average
\begin{equation}\label{ge}
\langle f(\lambda) \rangle =\frac{1}{|\cG|}\sum_{\sigma \in \cG}\sigma(f(\lambda))\, .
\end{equation}
In order to obtain a primitive function $f(\lambda,\gamma_1,\gamma_2)$ we define the automorphic function
\begin{equation}\label{fhat}
\hat{f}(\lambda,\gamma_1)=\langle \frac{1}{(\lambda-\gamma_1)^{|\cG _{\gamma_1}|}}\rangle=
\frac{1}{|\cG|}\sum_{\sigma \in \cG}\frac{1}{(\sigma^{-1}(\lambda)-\gamma_1)^{|\cG _{\gamma_1}|}}
\end{equation}
with poles of multiplicity $|\cG _{\gamma_1}|$ 
at points of the orbit $\cG (\gamma_1)$ and then $f(\lambda,\gamma_1 ,\gamma_2)=\hat{f}(\lambda,\gamma_1)-\hat{f}(\gamma_2,\gamma_1)$.
It is essential that the order of the pole in (\ref{fhat}) has been chosen equal to the order of the fixed point 
$\gamma_1$. If the order is less than $|\cG_{\gamma_1}|$, then the group average is a constant function, i.e. it does not depend on $\lambda$.

For completeness, in the Appendix we give an account of all finite groups (\ref{fgroups})
of fractional-linear transformations, their orbits 
and corresponding primitive automorphic functions.

\subsection{Automorphisms of semi-simple Lie algebras}

The structure of the automorphisms groups of semi-simple Lie algebras over $\bbbc$
is comprehensively studied (see for example the book of Jacobson \cite{jacob}).
In this section we list some results which will be used in the rest of 
the text.

  Let $\cA$ be a finite or infinite dimensional Lie algebra over any field (or ring). We denote by $\aut \cA$ the group of all automorphisms of $\cA$. 
Let $G\subset \aut\cA$
be a subgroup and $\cA _G$ be a subset of all elements of $\cA$ which are invariant with respect to all transformations of $G$, i.e.
\[ \cA _G=\{ a\in\cA\, |\, \phi (a)=a\, ,\, \forall \phi\in G\} \, .\]

\begin{lemma}\label{A_G} $\cA_G$ is a subalgebra of $\cA$. \end{lemma}

This lemma is obvious (it follows immediately from the automorphism definition),
but important for our further applications. All classical semi-simple Lie algebras  
can be extracted in such a way from the algebra of matrices with zero trace.
For example, the map $\phi_t (a)=-a^{tr}$, where $a^{tr}$ stands for the transpose matrix, is an automorphism of the Lie algebra $sl(N,\bbbc)$ of
square $N\times N$ matrices. The invariant subalgebra in this case is $so(N,\bbbc)$, i.e. the algebra of skew-symmetric matrices.

>From now on we assume that $\cA$ is a finite dimensional semi-simple Lie algebra over
$\bbbc$. The group $ \aut \cA$ is a Lie group. It is generated by inner automorphisms of the form  $\phi_{in}=e^{ad_a}\, ,a\in \cA$ and outer automorphisms $\phi_{out}$, induced by automorphisms (symmetries) of the Dynkin diagram of $\cA$. Any automorphism $\phi\in  \aut \cA$ can be uniquely represented as a composition $\phi_{in}\cdot \phi_{out}$. Inner automorphisms form a Lie subgroup $\aut\!_0 \cA$ of the group $\aut \cA$. The subgroup $\aut\!_0 \cA$ is normal and a connected component of the identity of the group of all automorphisms. The algebras $A_n\, , (n>1)\, ,\ D_n\, ,(n>4)$ and $E_6$ have subgroups
of outer automorphisms of order two, the algebra $D_4$ has the group $\aut \cA /\aut\!_0\cA \cong S_3$, i.e. the group of permutations of three elements, of order six and isomorphic to $\bbbd_3$. Other semi-simple Lie algebras do not admit outer automorphisms.

The description of the group of automorphisms can be given in explicit form. For example in the case of the algebra $sl(N,\bbbc)$ we have \cite{jacob}: 

\begin{theorem}\label{jacobth} The group of automorphisms of the Lie algebra of $2\times 2$ 
 matrices of zero trace is a set of mappings $a\to QaQ^{-1}$. The group of automorphisms 
of the Lie algebra of $N\times N\, ,N>2$, matrices of trace $0$ is a set of mappings 
$a\to QaQ^{-1}$ and  $a\to - Ha^{tr}H^{-1}$. Where $Q,H\in GL(N,\bbbc)$.
\end{theorem}

Explicit descriptions of the groups of automorphisms for other 
semi-simple algebras can be found in \cite{jacob}. In this paper we focus 
on the study of Lie subalgebras related to $sl(N,\bbbc)$.

\subsection{Automorphisms of Lie algebras over rings of rational functions. Automorphic Lie algebras}

A straightforward application of Lemma \ref{A_G} to finite dimensional semi-simple Lie algebras does not lead to interesting results. Indeed, if we wish the invariant subalgebra $\cA_G$ to be semi-simple we are coming back to the famous list of the Cartan classification and nothing new can be found on this way. Infinite dimensional Lie algebra with elements depending on a complex parameter $\lambda$ may have a richer group of automorphisms and Lemma \ref{A_G} provides a tool to construct subalgebras of infinite dimensional Lie algebras in the spirit of the theory of automorphic functions \cite{klein}, \cite{ford}.

Let $\Gamma =\{\gamma_1\, ,\ldots\, ,\gamma_N\}$ be a finite set of points $\gamma_k\in\hat{\bbbc}\simeq \cp1$. The linear space of all rational functions of a complex variable $\lambda\in\bbbc$ which may have poles of any finite order at points of $\Gamma$ and no other singularities in $\hat{\bbbc}$, equipped with usual multiplication, form a ring $\cR (\Gamma)$  and $\bbbc\subset \cR (\Gamma)$. The ring
$\cR (\Gamma)$, as a linear space of functions over $\bbbc$, is infinite dimensional.  Let $\cA$ be a finite dimensional semi--simple Lie algebra over $\bbbc$. We define
\beq
\cA(\Gamma)=\{\sum_{k}f_{k}(\lambda)\,e_{k}\,\,|\,\, f_{k}\in \cR(\Gamma)\, ,\,\, e_{k}\in\cA\,\} \, ,
\eeq
with standard commutator
\beq
\left[\sum_{k}f_{k}(\lambda)e_{k}\,,\,\sum_{s}g_{s}(\lambda)e_{s}\right]=\sum_{k,s}f_{k}(\lambda)g_{s}(\lambda)\left[e_{k}\, ,\,e_{s}\right] \, .
\eeq
The algebra $\cA(\Gamma)$ is an infinite dimensional Lie algebra over $\bbbc$. The group of automorphisms of $\cA(\Gamma)$, $\aut \cA(\Gamma)$, may be richer then $\aut \cA$; indeed, let $\Gamma$ be an orbit or a finite union of orbits of a finite group $\cG$ of fractional-linear transformations, then  the ring $\cR (\Gamma)$ has a nontrivial
group of automorphisms $\aut \cR (\Gamma) \cong \cG$. The group $\aut \cR (\Gamma)$ is the group of all
automorphisms of the ring which do not move the base field of constants (i.e. $\bbbc$). 
Automorphisms of the ring induce automorphisms of the algebra $\cA (\Gamma)$.
The direct product of the groups $\aut \cR (\Gamma)\times \aut \cA$ is a group
of automorphisms of  $\cA (\Gamma)$. It can be generalised to a semi-direct product, if there is a nontrivial homomorphism of $\aut \cR (\Gamma)$ in the group $\aut (\aut \cA)$ (an example will be given in Section \ref{semidirect}). In the rest of the article we assume that the set $\Gamma$ is an orbit or a union of a finite number of orbits of a finite group $\cG$ of fractional-linear transformations. 

For any group $H$ and two monomorphisms $\tau:H\to A$ and $\psi:H\to B$, the {\sl diagonal}
subgroup of the direct product $\tau(H)\times \psi(H)$ is defined as
\[ \mbox{diag}\, (\tau(H)\times \psi(H))=\{\big(\tau(h),\psi(h)\big)\, |\, h\in H\}\, .\]

Let $A$ and $B$ be two groups and $G$ be a subgroup of the direct product $G\subset A\times B$.
Each element $g\in G$ is a pair $g=(\alpha \, ,\beta)$, where $\alpha\in A$ and $\beta\in B$.
There are two natural projections $\pi_1, \pi_2$ on the first and the second components of the pair
\[ \pi_1 (g)=\alpha\, ,\qquad \pi_2 (g)=\beta \, .\]

\begin{theorem}\label{th1} Let $G\subset A\times B$ be a subgroup of the direct product of two groups $A,B$, and let
\[ U_{1}=G\cap (A\times id)\, ,\qquad U_{2}=G\cap (id\times B)\, ,\qquad K=U_{1}\cdot U_{2}\, .\]
 Then: 
\begin{enumerate}
\item $U_{1},U_{2}$ and $K$ are normal subgroups of $G$.
\item $\pi_i(U_{i})$ is a normal subgroup of $\pi_i(G)$, $i=1,\,2$.
\item There are two isomorphisms 
\[ \psi_1: G/K \to \pi_1(G)/\pi_1 (U_1)\, ,\quad \psi_2: G/K \to \pi_2(G)/\pi_2 (U_2)\, .\] 
\item 
$ G/K\cong \mbox{diag\, }(\psi_1(G/K)\times \psi_2(G/K))\, .$
\end{enumerate}
\end{theorem}

The proof of the theorem becomes obvious if we represent it in terms of two commutative diagrams ($i=1,\,2$) with exact horizontal and vertical sequences of  group homomorphisms:

\begin{diagram}
  &    & &    & &    &id &    &    \\
  &    & &    & &    &\dTo&   &    \\
id&\rTo&K&\rTo&G&\rTo&G/K&\rTo&id\\
  &    & \dTo_{\pi_i} &&\dTo_{\pi_i}&&\dTo_{\psi_i}&    &\\
id&\rTo&\pi_i (U_{i})&\rTo&\pi_i (G)&\rTo&\pi_i (G)/\pi_i (U_{i})&\rTo &id\\
  &    & \dTo&&\dTo& &\dTo&   &    &\\
  &    &  id &&id &  & id &   &    &  
\end{diagram}

\begin{df} Let $G\subset \aut \cA(\Gamma)$, we call the Lie 
algebra $\cA_G (\Gamma)$ \emph{automorphic}, if its elements $a\in\cA_G (\Gamma)$ 
are invariant $g(a)=a$ with respect to all automorphisms $g\in G$. Group $G$ is called the \emph{reduction group}.
\end{df}

The set $\cA_G (\Gamma)=\{a\in \cA (\Gamma)\, |\, g(a)=a,\, \forall g\in G\}$ is a subalgebra of $\cA (\Gamma)$ (Lemma \ref{A_G}).

Like automorphic functions, automorphic subalgebras of $\cA (\Gamma)$ can 
be constructed (in the case of a finite group $G$) using the group average.
For any element $a\in \cA (\Gamma)$ we define (compare with (\ref{ge}))
\begin{equation}\label{graverage0}
 \langle a \rangle _G=\frac{1}{|G|}\sum_{g\in G} g(a) \, .
\end{equation}
The group average is a liner operator in the linear space $\cA (\Gamma)$ over $\bbbc$,
moreover, it is a projector, since $\langle\langle a \rangle _G\rangle _G=\langle a \rangle _G$ for any element $a\in \cA (\Gamma)$.

If the group $G$ has a normal subgroup $N\subset G$ then we can perform the average in two stages: first we take the average 
over the normal subgroup $\bar{a}= \langle a \rangle _N $ and then take the average over the factor group $\langle \bar{a} \rangle _{G/N}$
\[  \langle a \rangle _G= \langle  \langle a \rangle _N \rangle _{G/N}\, . \]
Let $[g]$ be a co-set in $G/N$ and $\hat{g}\in [g]$ be one representative from the co-set, then the average $\langle \bar{a} \rangle _{G/N}$ is defined as
\[  \langle \bar{a} \rangle _{G/N}=\frac{|N|}{|G|}\sum_{[\hat{g}]\in G/N} \hat{g}(\bar{a}) \, .     \]
This definition is well posed since $\hat{g}(\bar{a})$ is constant on each co-set $[g]$, i.e. the result does not depend on the choice of a representative.

If $G\subset \aut \cR(\Gamma) \times \aut  \cA$, and it has nontrivial normal subgroups
$U_{1},U_{2}$  (in the notation of Theorem \ref{th1}) then
\begin{equation}\label{factoraverage}
\cA_G(\Gamma)=\langle \cA(\Gamma) \rangle _G=\langle \langle \langle \cA(\Gamma) \rangle _{U_{1}} \rangle _{U_{2}} \rangle _{G/K}=
\langle \langle \langle \cA(\Gamma) \rangle _{U_{2}} \rangle _{U_{1}} \rangle _{G/K}\, .
\end{equation}

The normal subgroup $U_{1}$ of a reduction group $G$ 
consists of all elements of the form $(\sigma,id)$,  it corresponds to
fractional-linear transformations of the complex variable $\lambda$, and identical
transformation of the algebra $\cA$.  The normal subgroup $U_{2}$ consists of all elements of the 
form $(id,\phi)$, i.e. automorphisms of $\cA$ and identical transformation of 
 the variable $\lambda$. The factor group $G/K$, if it is nontrivial, corresponds to
 simultaneous automorphisms of the ring $\cR(\Gamma)$ and the algebra 
$\cA$. 

Averaging $\cA (\Gamma)$ over $U_{2}$ is equivalent to a replacement of the algebra
$\cA$ by $\cA_{\pi_2(U_{2})}$ (Lemma \ref{A_G}). Thus, without any loss of generality, we can start from a smaller algebra $\cA_{\pi_2 (U_{2})}$ and respectively a smaller reduction  group 
$\hat{G}\cong G/U_{2}$. 

Averaging over $U_{1}$ affects only the ring $\cR (\Gamma)$. As the result, 
 we receive a subring $\cR_{\pi_1(U_{1})}(\Gamma)\subset \cR (\Gamma)$ of  $\pi_1(U_{1})$-automorphic functions with poles at $\Gamma$. It follows from the Theorem 
\ref{primfunth} that any element of $\cR_{\pi_1(U_{1})}(\Gamma)$ can be expressed as 
a rational function of a primitive $\pi_1(U_{1})$-automorphic function. Taking a 
primitive automorphic function instead of $\lambda$, we reduce then the problem to a simpler one
(with a trivial subgroup $U_{1}$), without any loss of generality.

Thus, the most interesting case corresponds to simultaneous transformations and from the very beginning we can assume the subgroup $K=U_{1}\cdot U_{2}$ to be trivial, without any loss of generality. If $K$ is trivial then   $G\cong \cG =\pi_1(G)\simeq\pi_2(G)$ (Theorem \ref{primfunth}). If 
$G$ is finite, it should be isomorphic to one of the finite groups of fractional-linear transformations (\ref{fgroups}). Thus, the reduction group
 $G =\mbox{diag}\, (\cG,\psi(\cG))$ where  $\psi :\cG\to \aut{\cA}$ is a monomorphism of a finite group of fractional-linear transformations $\cG$ into the group of automorphisms of Lie algebra $\cA$. 

The above construction can be generalised to the case in which the elements of $\aut{\cA}$ are $\lambda$ dependent. In this case, the composition law for the elements of the reduction group is similar to the one for a semi-direct product of groups. A nontrivial example of such generalisation and the corresponding automorphic Lie algebra will be discussed in section \ref{semidirect}.

\subsection{Quasigraded structure}

Following I. M. Krichever and S. P. Novikov \cite{kn87}  we define a \emph{quasigraded structure} for infinite dimensional Lie algebras.

\begin{df}\label{def:qg} An infinite dimensional Lie algebra $\cL$ is called \emph{quasigraded}, if it admits a decomposition as a vector space in a direct sum of subspaces
\beq\label{decomposition}
\cL=\bigoplus_{n\in\bbbz}\cL^{n}\, 
\eeq
and there exist two non negative integer constants $p$ and $q$ such that
\beq\label{qgrading}
[\cL^{n},\cL^{m}]\subseteq\bigoplus_{-q\leq k\leq p}\cL^{n+m+k}\quad\forall\, n,\, m\in\bbbz\, .
\eeq
For $p=q=0$ the algebra $\cL$ is graded. Elements of $\cL^n$ are called homogeneous elements of degree $n$. The decomposition (\ref{decomposition}) with the property
 (\ref{qgrading}) is called a \emph{quasigraded structure} of $\cL$.
\end{df}

Without loss of generality we can assume $q=0$. Indeed, by a simple shift in the enumeration we can always set $q=0$.  Quasigraded algebras with $p=1, q=0$ share one important property with graded algebras ($p=q=0$), namely they can be decomposed (splitted)
into a sum of two subalgebras \[ \cL=\cL_+\bigoplus \cL_-\] where
\[ \cL_+ =  \bigoplus_{n\ge 0}\cL^{n}\, ,\quad \cL_- =  \bigoplus_{n<0}\cL^{n}\, .
\]
Indeed, it follows from (\ref{qgrading}) that commutators of elements from subspaces  with negative indexes belong to $\cL_-$ since $n+m+1<0$ if $n<0$ and $m<0$. Commutators of elements from $\cL_+$ obviously belong to $\cL_+$. 
If $q=0$ and $p>1$ then $\cL_-$ is not necessarily a closed subalgebra, but $\cL_+$ is.

\section{Explicit construction of  automorphic Lie algebras}

To construct an automorphic Lie algebra we consider the following:
\begin{enumerate}
\item a finite group of fractional-linear transformations $\cG$,
\item a finite dimensional semi-simple Lie algebra $\cA$ over $\bbbc$,
\item a monomorphism $\psi :\cG\to\aut \cA$.
\end{enumerate}

For a given $\cG,\cA$ and $\psi$, automorphic Lie algebras depend on the choice of a $\cG$-invariant set $\Gamma$, which is a union of a finite number of orbits $\Gamma=\cup_{k=1}^M \cG (\gamma_k)$. Similar to the theory of automorphic functions (Theorem \ref{primfunth}), for any two orbits $\cG(\gamma_1), \cG(\gamma_2)$, there is a uniquely defined {\sl primitive} automorphic Lie algebra $\cA_\cG (\gamma_1,\gamma_2)$, whose elements may have poles at points in
$\cG (\gamma_1)\cup \cG(\gamma_2) $ and do not have any other singularities.  Algebra $\cA_\cG (\gamma_1,\gamma_2)$ is \emph{quasigraded} (see Definition \ref{def:qg})  and its structure constants can be written explicitly. Structure constants of any other $\cG$-automorphic Lie algebra can be explicitly expressed in terms of the structure constants of $\cA_\cG (\gamma_1,\gamma_2)$. In general, algebra $\cA_\cG (\gamma_1,\gamma_2)$ can be decomposed
in a direct sum of three linear spaces
\begin{equation}\label{decomp0} 
\cA_\cG (\gamma_1,\gamma_2)=\cA_\cG (\gamma_1)\bigoplus\cA_\cG^0 \bigoplus
\cA_\cG (\gamma_2)\, ,
\end{equation}
such that elements of  $\cA_\cG (\gamma)$ may have poles at the points of the orbit 
$\cG(\gamma)$ and are regular elsewhere and elements of a finite dimensional linear space $\cA_\cG^0$  are constants, i.e. they do not depend on $\lambda$. Often the subspace $\cA_\cG^0$ is  empty, then $\cA_\cG (\gamma_1)$ and $\cA_\cG (\gamma_2)$ are subalgebras. In all cases studied we have found a subalgebra $\hat{\cA}_\cG (\gamma_1,\gamma_2)\subseteq \cA_\cG (\gamma_1,\gamma_2)$ which can be decomposed as a linear space in a direct sum 
\begin{equation}\label{decomp} 
\hat{\cA}_\cG (\gamma_1,\gamma_2)=\hat{\cA}_\cG (\gamma_1)\bigoplus
\hat{\cA}_\cG (\gamma_2)\, ,
\end{equation}
such that $\hat{\cA}_\cG (\gamma_1)$ and $
\hat{\cA}_\cG (\gamma_2)$ are subalgebras whose elements may have poles at the orbits $\cG(\gamma_1)$ and $\cG(\gamma_2)$ respectively and are regular elsewhere.
 
\subsection{Simple example $\cG =\bbbd_N \, ,\, \cA =sl(2,\bbbc)$.}

The action of the dihedral group $\bbbd _N$ on the complex plane can be generated 
by two transformations $\sigma_s (\lambda)=\Omega \lambda$, with $\Omega=\exp(2i\pi/N)$ and $\sigma_t (\lambda)=\lambda^{-1}$ (see details in the Appendix). It follows from Theorem \ref{jacobth} that all automorphisms $\aut sl(2,\bbbc)$ are inner and can be represented in the form $\phi (a)=QaQ^{-1}$ where $Q\in GL(2,\bbbc)$. A monomorphism $\psi :\bbbd_N \to \aut sl(2,\bbbc)$ is nothing but a faithful projective representation of $\bbbd_N$ and it is sufficient to define it on the generators of the group. Let $Q_s$ and $ Q_t$ correspond to $\sigma_s$ and $ \sigma _t$, respectively. Two projective representations $Q_s, Q_t$ and $\hat{Q}_s, \hat{Q}_t$ are equivalent
if there exist $W\in GL(2,\bbbc)$ and $c_s,c_t\in\bbbc$ such that $WQ_s W^{-1}=c_s \hat{Q}_s$ and
$WQ_t W^{-1}=c_t \hat{Q}_t$. 

In the simplest case $\cG =\bbbd_2\cong \bbbz_2\times \bbbz_2$ there is only one class of faithful projective representations which is equivalent to the choice
\begin{equation}\label{QsQt}
 Q_s =\left(  \begin{array}{rr} 1&0\\0&-1\end{array}\right) \, ,\quad  
Q_t=\left(  \begin{array}{rr} 0&1\\1&0\end{array}\right)\, .
\end{equation} 
 Thus the reduction group $\bbbd_2$ can be generated by two transformations
\[ g_s\,:\,\,\,a(\lambda)\to Q_s a(-\lambda) Q_s^{-1}\, ,\quad g_t\,:\,\,\,a(\lambda)\to Q_t a(\lambda^{-1}) Q_t^{-1}\, ,
\quad a(\lambda)\in \cA (\Gamma) \]
and the group average is
\[ \langle a(\lambda)\rangle _{\bbbd_2}=\frac{1}{4}(a(\lambda)+Q_s a(-\lambda) Q_s^{-1}+
Q_t a(\lambda^{-1}) Q_t^{-1}+Q_t Q_s a(-\lambda^{-1}) Q_s^{-1}Q_t^{-1})\, .\]

In $\cA =sl(2,\bbbc)$ we take the standard basis 
\begin{equation}\label{xyhbasis} h=\left(  \begin{array}{rr} 1&0\\0&-1\end{array}\right) \, ,\quad
x=\left(  \begin{array}{rr} 0&1\\0&0\end{array}\right) \, ,\quad
y=\left(  \begin{array}{rr} 0&0\\1&0\end{array}\right) \, ,\quad
\end{equation}
with commutation relations
\[ [x,y]=h\, ,\quad  [h,x]=2x\, ,\quad [h,y]=-2y\, .\]

Let $\gamma\in \hat{\bbbc}$ be a generic point, i.e. $\gamma\not\in\{0,\infty,\pm 1,\pm i\}$ and $|\cG_{\gamma}|=1$,
then
\begin{eqnarray} \label{xg}
x_\gamma(\lambda ) &=&\langle \frac{x}{\lambda -\gamma}\rangle _{\bbbd_2}=\left( 
\begin{array}{cc} 0& \frac{\lambda}{2(\lambda ^2-\gamma^2)}\\
\frac{\lambda}{2(1-\lambda ^2\gamma^2)} & 0\end{array}\right) \\
\label{yg} y_\gamma (\lambda ) &=&\langle \frac{y}{\lambda -\gamma}\rangle _{\bbbd_2}=\left( 
\begin{array}{cc} 0& \frac{\lambda}{2(1-\lambda ^2\gamma^2)} \\
\frac{\lambda}{2(\lambda ^2-\gamma^2)}& 0\end{array}\right) \\
\label{hg} h_\gamma (\lambda )&=&\langle \frac{h}{\lambda -\gamma}\rangle _{\bbbd_2}=
\frac{\gamma (1-\lambda^4)}{2 (\lambda ^2-\gamma^2)(1-\lambda ^2\gamma^2)}
\left( 
\begin{array}{cc} 1&0\\
0 & -1\end{array}\right) 
\end{eqnarray}

We shall denote $sl_{\bbbd_2} (2,\bbbc;\gamma )$ the infinite dimensional 
Lie algebra  of all $\bbbd_2$-automorphic traceless $2\times 2$ matrices whose entries are rational functions in $\lambda$ with poles at $\bbbd_2 (\gamma)$ and with no other singularities.

\begin{prop}\label{propbasis} Let $\mu\in \bbbc \setminus\{\pm \gamma,\pm \gamma^{-1}\}$. The set 
\begin{equation}\label{basis+}\begin{array}{l}
x^n_{\gamma\mu}=4 x_\gamma(\lambda) (f_{\bbbd_2}(\lambda, \gamma, \mu))^{n}\\
y^n_{\gamma\mu}=4 y_\gamma(\lambda) (f_{\bbbd_2}(\lambda, \gamma, \mu))^{n}\\
h^n_{\gamma\mu}=4 h_\gamma(\lambda) (f_{\bbbd_2}(\lambda, \gamma, \mu))^{n}
\end{array}\, ,\qquad n=0,1,2,\ldots 
\end{equation}
is a basis in $sl_{\bbbd_2} (2,\bbbc;\gamma )$. Here $f_{\bbbd_2}(\lambda,\gamma,\mu)$ is a primitive automorphic function defined as
\begin{equation}\label{alpha} f_{\bbbd_2}(\lambda, \gamma, \mu)=\alpha 
\frac{(\lambda^2-\mu^2)(1-\mu^2\lambda^2)}{(\lambda^2-\gamma^2)(1-\gamma^2\lambda^2)}\, , \quad \alpha = \frac{2\gamma(\gamma^4-1)}{(\mu^2-\gamma^2)(1-\mu^2\gamma^2)} \, .
\end{equation}
\end{prop}

In (\ref{alpha}) we have chosen the constant $\alpha $ to make $res_{\lambda=\gamma} f_{\bbbd_2}(\lambda,\gamma,\mu)=1$.

{\bf Proof} We prove the proposition by induction. Let $a(\lambda)\in sl_{\bbbd_2} (2,\bbbc;\gamma )$. If $a(\lambda)=a_{0}$ does not have a singularity at $\lambda=\gamma$, then $a_{0}=0$. Indeed, in this case $a_{0}$ does not have singularities at all and therefore it is a constant matrix. It follows from $g_{s}(a_{0})=a_{0}$ and $g_{t}(a_{0})=a_{0}$ that $a_{0}$ commutes with $Q_{s}$ and $Q_{t}$, therefore $a_0$ has to be proportional to the unit matrix. From $trace(a_{0})=0$ follows that $a_0=0$. Suppose $a(\lambda)$ has a pole of order $n>0$
at $\lambda=\gamma$, then near the singularity it can be represented as $a(\lambda)=a_0 (\lambda-\gamma)^{-n}+\hat{a}(\lambda )$
where $a_0$ is a constant matrix, $ \hat{a}(\lambda )$ may have a pole at $\lambda=\gamma$ of order $m<n$. In the basis (\ref{xyhbasis})  $a_0=c_1 x+c_2 y+c_3 h,\ c_i\in\bbbc$. If
\[ b(\lambda)=a(\lambda)-4 (c_1 x_\gamma (\lambda)+c_2 y_\gamma (\lambda)+c_3 h_\gamma (\lambda)) f_{\bbbd_2}^{n-1}(\lambda, \gamma, \mu)\in sl_{\bbbd_2} (2,\bbbc,\gamma )\]
is singular at 
$\lambda=\gamma$ then the order of its pole is less or equal to $n-1$ and this complete the induction step. $\square$

\begin{prop} 
 Elements $x_\gamma(\lambda ),y_\gamma(\lambda ),h_\gamma(\lambda )$ generates a $\bbbd_2$-automorphic Lie algebra $sl_{\bbbd_2} (2,\bbbc;\gamma )$. The algebra $sl_{\bbbd_2} (2,\bbbc;\gamma ) $ is quasigraded, its quasigraded structure 
\[sl_{\bbbd_2} (2,\bbbc;\gamma )= \bigoplus_{n\in\bbbz}\cL_{\gamma}^{n}(\mu)\, ,\qquad 
[\cL_{\gamma}^{n}(\mu),\cL_{\gamma}^{m}(\mu)]\subseteq \cL_{\gamma}^{n+m+1}(\mu)\bigoplus \cL_{\gamma}^{n+m}(\mu)
\]
depends on a complex parameter $\mu$ and $\cL_\gamma^n(\mu) =\mbox{Span}_\bbbc\,(x_{\gamma\mu}^n,y_{\gamma\mu}^n,h_{\gamma\mu}^n)$. 
\end{prop}

{\bf Proof} Indeed, by direct calculation we find that
\begin{eqnarray}\nonumber
&&\left[ x^n_{\gamma\mu},y^m_{\gamma\mu}\right] = h_{\gamma\mu}^{n+m+1}+a_{\gamma\mu}  h_{\gamma\mu}^{n+m} \\ \label{commrel}
&&\left[ h ^n_{\gamma \mu}, x^m_{\gamma \mu}\right]  = 2 x_{\gamma\mu}^{n+m+1}+b_{\gamma\mu}x_{\gamma\mu}^{n+m}-c_{\gamma\mu}y_{\gamma\mu}^{n+m}\, ,\qquad n,m=0,1,2,\ldots \\ \nonumber
&&\left[ h^n_{\gamma\mu},y^m_{\gamma\mu}\right] =-2 y_{\gamma\mu}^{n+m+1}-b_{\gamma\mu}y_{\gamma\mu}^{n+m}+c_{\gamma\mu}x_{\gamma\mu}^{n+m}
\end{eqnarray}
where
\[
a_{\gamma\mu}=\frac{2\mu^2(1-\gamma^4)}{\gamma(\mu^2-\gamma^2)(1-\mu^2\gamma^2)},\quad
b_{\gamma\mu}=\frac{4\gamma (1+\mu^4-4\mu^2\gamma^2+\gamma^4+\gamma^4\mu^4)}{(1-\gamma^4)(\mu^2-\gamma^2)(1-\mu^2\gamma^2)},\quad 
 c_{\gamma\mu}=\frac{8\gamma}{1-\gamma^4}\, .
\]
Thus, any element of the basis (\ref{basis+}) can be generated by the set (\ref{xg})--(\ref{hg}). It follows from (\ref{commrel}) that $q=0,p=1$ (see (\ref{qgrading})). $\square$

The quasigraded structure of  $sl_{\bbbd_2} (2,\bbbc;\gamma ) $, i.e its decomposition in a direct sum of linear subspaces $\cL_\gamma^n(\mu)$, depends on a complex parameter $\mu$. This parameter  determines the zeros of the primitive automorphic function
$f_{\bbbd_2}(\lambda,\gamma,\mu)$. Taking into account the fact that $f_{\bbbd_2}(\lambda,\gamma,\nu)=
f_{\bbbd_2}(\lambda,\gamma,\mu)-f_{\bbbd_2}(\nu,\gamma,\mu)$, we see that the corresponding bases $\{x^n_{\gamma\mu},y^n_{\gamma\mu},h^n_{\gamma\mu}  \}_{n\in \bbbz_+}$ and $\{x^n_{\gamma\nu},y^n_{\gamma\nu},h^n_{\gamma\nu}  \}_{n\in \bbbz_+}$ are related by a simple invertible triangular transformation
\begin{equation} \label{treug}
x_{\gamma \nu}^n=\sum_{k=0} (-1)^k {n \choose k} (f_{\bbbd_2}(\nu,\gamma,\mu))^{k}\,x_{\gamma\mu}^{n-k} 
\end{equation}
(same for $y_{\gamma \nu}^n$ and $h_{\gamma \nu}^n$), where ${n \choose k}$ are binomial coefficients. For positive $n$ the sum (\ref{treug}) is finite, since all ${n \choose k}$ vanish as $k>n$.

The set $\{ x_{\gamma\mu}^n,y_{\gamma\mu}^n,h_{\gamma\mu}^n\}$ is naturally defined for negative integers $n\in \bbbz_-$. 

\begin{prop}\label{negbasis} Elements  $x_{\gamma\mu}^{-1},y_{\gamma\mu}^{-1},h_{\gamma\mu}^{-1}$ generate a  
$\bbbd_2$-automorphic Lie algebra $sl_{\bbbd_2} (2,\bbbc;\mu)$. The set $\{ x_{\gamma\mu}^n,y_{\gamma\mu}^n,h_{\gamma\mu}^n\, |\, n\in\bbbz_- \}$ is a basis 
in $sl_{\bbbd_2} (2,\bbbc,\mu)$.
\end{prop}
 
{\bf Proof} For negative $n$, automorphic elements  $x_{\gamma\mu}^n,y_{\gamma\mu}^n,h_{\gamma\mu}^n$ have poles at points $\cG(\mu)$ and do not have other singularities, therefore  $x_{\gamma\mu}^n,y_{\gamma\mu}^n,h_{\gamma\mu}^n\in sl_{\bbbd_2} (2,\bbbc;\mu)$. The proof that $\{ x_{\gamma\mu}^n,y_{\gamma\mu}^n,h_{\gamma\mu}^n\, |\, n\in\bbbz_-  \}$ form a basis in $sl_{\bbbd_2} (2,\bbbc;\mu)$ is similar to Proposition \ref{propbasis}. $\square$

Thus, with any two orbits $\bbbd_2(\gamma)$ and $\bbbd_2(\mu)$ we associate two uniquely defined subalgebras $ sl_{\bbbd_2} (2,\bbbc,\gamma)$ and $sl_{\bbbd_2} (2,\bbbc,\mu)$ of the infinite dimensional Lie algebra 
\[ sl_{\bbbd_2} (2,\bbbc;\gamma,\mu) =sl_{\bbbd_2} (2,\bbbc,\gamma)\bigoplus sl_{\bbbd_2} (2,\bbbc,\mu)\, .\]
The set (\ref{basis+}) with $n\in \bbbz$ is a basis in  $ sl_{\bbbd_2} (2,\bbbc;\gamma,\mu)$
with commutation relations (\ref{commrel}).
$sl_{\bbbd_2} (2,\bbbc;\gamma,\mu)$ has a uniquely defined quasigraded structure corresponding to a primitive automorphic function $f_{\bbbd_2}(\lambda,\gamma,\mu)$. 
Quasigraded automorphic algebras corresponding to different orbits are not isomorphic, i.e. elements of one algebra cannot be represented by finite linear combination of the basis elements of the other algebra with complex constant coefficients.

In the above construction, the point $\mu$ could be a generic point or belong to one of the degenerated orbits. Having generators and structure constants for algebra $sl_{\bbbd_2} (2,\bbbc,\gamma) $ we can easily find generators and corresponding structure constants for $sl_{\bbbd_2} (2,\bbbc,\mu) $. 
Taking, for example, $\mu=0$ we find generators
\begin{equation} \label{hatxyh0}
\hat{x}_0=4 x_\gamma (f_{\bbbd_2}(\lambda,\gamma,0))^{-1}
\, ,\quad \hat{y}_0=4 y_\gamma (f_{\bbbd_2}(\lambda,\gamma,0))^{-1}
\, ,\quad \hat{h}_0=4 h_\gamma (f_{\bbbd_2}(\lambda,\gamma,0))^{-1}
\end{equation} 
for $sl_{\bbbd_2} (2,\bbbc,0)$. The set $\{ \hat{x}_0^n=\hat{x}_0 (f_{\bbbd_2}(\lambda,\gamma,0))^{-n},\hat{y}_0^n=\hat{y}_0  (f_{\bbbd_2}(\lambda,\gamma,0))^{-n},\hat{h}_0^n=\hat{h}_0  (f_{\bbbd_2}(\lambda,\gamma,0))^{-n}|n\in\bbbz_+ \}$ can be taken as a basis (compare with Proposition \ref{negbasis}). The structure constants 
in this basis follows immediately from (\ref{commrel}).

The generators of $sl_{\bbbd_2} (2,\bbbc,0)$ can also be found directly, by taking the group average
\begin{eqnarray} \nonumber
&& x_0(\lambda ) =\langle \frac{x}{\lambda}\rangle _{\bbbd_2}=\frac{1}{2}\left(\begin{array}{cc} 0& \lambda^{-1}\\
\lambda & 0\end{array}\right)\, ,\\ \label{x0y0h0}
&&y_0 (\lambda ) =\langle \frac{y}{\lambda}\rangle _{\bbbd_2}=\frac{1}{2}\left(\begin{array}{cc} 0& \lambda \\
\lambda^{-1}& 0\end{array}\right)\, , \\ \nonumber
&&h_0 (\lambda )=\langle \frac{h}{\lambda^2}\rangle _{\bbbd_2}=
\frac{(1-\lambda^4)}{2 \lambda ^2}
\left( 
\begin{array}{cc} 1&0\\
0 & -1\end{array}\right) \, .
\end{eqnarray}
Generators (\ref{hatxyh0}) can be expressed in terms of (\ref{x0y0h0})
\[ 
\hat{x}_0=\frac{2\gamma}{1-\gamma^4}(x_0-\gamma^2 y_0)\, ,\quad 
\hat{y}_0=\frac{2\gamma}{1-\gamma^4}(y_0-\gamma^2 x_0)\, ,\quad
\hat{h}_0=\frac{8\gamma^2}{1-\gamma^4}h_0 \, .\]
In the basis $\{ x_0^n=x_0 J^{n},y_0^n=y_0 J^{n},h_0^n=\frac{1}{2}h_0 J^{n}\}_{n\in\bbbz_+} $ where $J=f_{\bbbd_2}(\lambda, 0, 1)=\frac{1}{2}(\lambda - \lambda^{-1})^2$, the commutation relations
of $sl_{\bbbd_2}(2,\bbbc,0)$ take a very simple form:
\[
[x_0^n,y_0^m]=h_0^{n+m}\, ,\quad [h_0^n,x_0^m]=x_0^{n+m+1}+x_0^{n+m}-y_0^{n+m}
\, ,\quad [h_0^n,y_0^m]=-y_0^{n+m+1}-y_0^{n+m}+x_0^{n+m} .
\] 

In the case $\cG \cong \bbbd_3$ the projective representation is
generated by $Q_t$ (\ref{QsQt}) and  $Q_{s}=\mbox{diag}\, (e^{\frac{2 \pi i}{3}},
e^{-\frac{2\pi i}{3}})$.
Using the group average one can find $sl_{\bbbd_3}(2,\bbbc,\gamma)$ algebra generators 
and then the basis in which the algebra has a quasigraded structure. It turns out that 
 algebra 
$sl_{\bbbd_3}(2,\bbbc,\gamma)$
is isomorphic to $sl_{\bbbd_2}(2,\bbbc,\mu)$ if $\gamma^3=\mu^2$. In particularly  
$sl_{\bbbd_3}(2,\bbbc,0)\cong sl_{\bbbd_2}(2,\bbbc,0)$. It is a general observation --
for any $N,M\ge 2$ and $\gamma\in\bbbc$ 
\[ 
sl_{\bbbd_N}(2,\bbbc,\gamma^M)\cong sl_{\bbbd_M}(2,\bbbc,\gamma^N)\, .
\]
For $N>2$ there is a choice of inequivalent irreducible representations of $\bbbd_N$. Automorphic Lie algebras corresponding to different representations proved to be isomorphic. This explains why integrable equations corresponding to $\bbbd_N$ reductions with different $N$ and non equivalent representations coincide \cite{lm1}.

\subsection{Automorphic Lie algebras with $\cG =\bbbd_N \, ,\, \cA =sl(3,\bbbc)$.}

Let the action of $\bbbd_N$ on the complex plane $\lambda$ be the same as in the previous section, i.e. generated by two fractional-linear transformations $\sigma_s (\lambda)=\Omega \lambda$ and $\sigma_t (\lambda)=\lambda^{-1}$ with $\Omega=\exp(2i\pi/N)$. 

It follows from Theorem \ref{jacobth} that automorphisms $\aut (sl(3,\bbbc))$ can be represented either in the form $a\to QaQ^{-1}$ or $a\to -Ha^{tr}H^{-1}$   where $Q,H\in GL(3,\bbbc)$. The first kind of automorphisms (with $Q$) form a normal subgroup of inner automorphisms $\aut _0 (sl(3,\bbbc))$, while automorphisms with $H$ correspond to outer automorphisms and $\aut ( sl(3,\bbbc))/\aut _0 (sl(3,\bbbc))\cong\bbbz_2$. There are two distinct ways to define a monomorphism $\psi: \bbbd_N \to \aut  (sl(3,\bbbc))$:
\begin{description}
\item[Case A] $\psi$ maps $\bbbd_N$ into the subgroup of inner automorphisms (similar to the previous section). In this case $\psi$ is nothing but a faithful projective representation of $\bbbd_N$. 
\item[Case B] The  other option is to use a normal subgroup decomposition
($id\to\bbbz_N\to\bbbd_N\to\bbbz_2\to id$). In this case $\psi$ maps the normal subgroup $\bbbz_N$ in $\aut _0 (sl(3,\bbbc))$, and its co-set into the co-set corresponding to outer automorphisms, so that the following commutative diagram is exact:
\begin{diagram}
  &    & id    &    &id         &    &id    &   & \\
  &    & \dTo  &    &\dTo  &    &\dTo   &   & \\
id&\rTo&\bbbz_N&\rTo&\bbbd_N    &\rTo&\bbbz_2&\rTo&id\\
  &    & \dTo_{\psi}  &    &\dTo_{\psi}&    &\dTo   &   & \\
id&\rTo&\aut _0 (sl(3,\bbbc))&\rTo&\aut(sl(3,\bbbc))&\rTo&\bbbz_2&\rTo&id
\end{diagram}
\end{description}
 
We shall study these two cases separately.

\subsubsection{Case A. Inner automorphisms representation.}

We shall see that in the case $\cA=sl(3,\bbbc)$ the reduction groups $\bbbd_2^A$ and $\bbbd_N^A, N>2$ yield non-isomorphic automorphic Lie algebras (the upper index stands for the case A). Let $\e_{ij}$ denotes a matrix with $1$ at the position $(i,j)$ and zeros elsewhere. Matrices $\e_{ij}\, ,\ i\ne j$ and $h_1=\e_{11}-\e_{22}\, ,\ h_2=\e_{22}-\e_{33}$ form a basis in $sl(3,\bbbc)$.

{\em Case A,} $G=\bbbd_2^A$: The action of the reduction group $G=\bbbd_2^A$
can be generated by transformations 
\begin{equation}\label{d2_3}
 g_s: a(\lambda)\to Q_sa(-\lambda)Q_s^{-1} \, ,\quad g_t: a(\lambda)\to Q_t a(1/\lambda)Q_t ^{-1}\, ,
\end{equation}
where $Q_s=\mbox{diag}\, (-1,1,-1)$ and $Q_t =\mbox{diag}\, (1,-1,-1)$. It is easy to check that $g_s^2=g_t^2=(g_s g_t)^2=id$.  If one ignores the $\lambda$ transformations, then (\ref{d2_3}) form a $\bbbd_2$ subgroup of inner automorphisms of algebra $sl(3,\bbbc)$. 

In order to fix a primitive automorphic Lie algebra, we need to fix two orbits of the reduction group on the complex plane $\lambda$. As in the previous section, the choice of the orbits is not very essential, since knowing the structure constants of the algebra for one choice of the orbits, we can easily reconstruct the structure constants for any other choice. We shall consider the
orbits $\{0,\infty\}$ and $\{1,-1\}$ and take the corresponding primitive automorphic function in the form
\[J=f_{\bbbd_2}(\lambda,0,1)=\lambda^2 +\lambda^{-2}-2\, .\]

Automorphic Lie algebra $sl_{\bbbd_2^A}(3,\bbbc; 0,1)$ is quasigraded ($[\cA_n,\cA_m]\subset \cA_{n+m}\bigoplus \cA_{n+m+1}\bigoplus \cA_{n+m+2}$), 
\[ sl_{\bbbd_2^A}(3,\bbbc; 0,1)=\bigoplus_{n\in \bbbz}\cA_n\,  ,\] 
where $\cA_n = J^{n}\cA_0$. It is sufficient to give a description of the linear space $\cA_0$ and commutation relations $[\cA_0,\cA_0]$. A basis in the eight dimensional space $\cA_0$ can be chosen as:
\begin{equation}\label{basis30} \begin{array}{ll}
x_1^0=\langle 2\e_{12}\lambda^{-1}\rangle_{\bbbd_2^A}=(\lambda^{-1}-\lambda)\e_{12}\, ,&
y_1^0=\langle 2\e_{21}\lambda^{-1}\rangle_{\bbbd_2^A}=(\lambda^{-1}-\lambda)\e_{21}\\
x_2^0=\langle 2\e_{23}\lambda^{-1}\rangle_{\bbbd_2^A}=(\lambda^{-1}+\lambda)\e_{23}\, ,&
y_2^0=\langle 2\e_{32}\lambda^{-1}\rangle_{\bbbd_2^A}=(\lambda^{-1}+\lambda)\e_{32}\\
x_3^0=[x_1^0,x_2^0]=(\lambda^{-2}-\lambda^2)\e_{13}\, ,&
y_3^0=[y_2^0,y_1^0]=(\lambda^{-2}-\lambda^2)\e_{31}\end{array}\end{equation}
\begin{equation}\label{basis30h}
 h_1^0=\e_{11}-\e_{22}\, ,\qquad h_2^0=\e_{22}-\e_{33}\, .\end{equation}
\begin{prop} The set 
\begin{equation}\label{basis3}
x_i^n=J^{n}x_i^0\, ,\quad y_i^n= J^{n }y_i^0\, ,\quad  h_j^n= J^{n }h_j^0\, ,\quad i\in\{1,2,3\},\quad j\in\{1,2\},\quad  n\in \bbbz
\end{equation} 
is a basis of the algebra  $sl_{\bbbd_2^A}(3,\bbbc; 0,1)$.
\end{prop}

The proof is similar to Propositions \ref{propbasis}, \ref{negbasis}. It is easy to compute all commutators between the basis elements of $sl_{\bbbd_2^A}(3,\bbbc; 0,1)$. For example
\[ [h^n_1, x^m_1]=2 x_1^{n+m}\, ,\quad  [x^n_1, y^m_1]=h^{n+m+1}_1-2 h_1^{n+m}\, ,\quad  
[x^n_3, y^m_3]=h^{n+m+2}_1+h^{n+m+2}_2-4 h_1^{n+m}-4 h_2^{n+m}\, .\]

Algebra $sl_{\bbbd_2^A}(3,\bbbc; 0,1)$ has a quasigraded subalgebra $\hat{sl}_{\bbbd_2^A}(3,\bbbc; 0,1)$,
which is a direct sum of two infinite dimensional subalgebras
\[ \hat{sl}_{\bbbd_2^A}(3,\bbbc; 0,1)=\hat{sl}_{\bbbd_2^A}(3,\bbbc; 0)\bigoplus \hat{sl}_{\bbbd_2^A}(3,\bbbc; 1)\]
As a basis in $\hat{sl}_{\bbbd_2^A}(3,\bbbc; 0,1)$  we can take a set $x_i^n, y_i^n$ defined in (\ref{basis30}), (\ref{basis3}) and 
\[
 \hat{h}_1^n=J^n [x_1^0,y_1^0]=(\lambda^2+\lambda^{-2}-2)J^n (\e_{11}-\e_{22})\, ,\quad \hat{h}_2^n=J^n[x_2^0,y_2^0]=(\lambda^2+\lambda^{-2}+2)J^n (\e_{22}-\e_{33})\, ,\qquad n\in\bbbz .\]
In this basis the non-vanishing commutation relations are
{\small\bd \begin{array}{lll}
\br\hat{h}_1^n,x_1^m]=2x_1^{n+m+1}, & [\hat{h}_1^n,y_1^m]=-2y_1^{n+m+1}, & [\hat{h}_1^n,x_2^m]=-x_2^{n+m+1}, \\
\br\hat{h}_1^n,y_2^m]=y_2^{n+m+1}, & [\hat{h}_1^n,x_3^m]=x_3^{n+m+1}, & [\hat{h}_1^n,y_3^m]=-y_3^{n+m+1}, \\
\br\hat{h}_2^n,x_1^m]=-x_1^{n+m+1}-4x_1^{n+m}, & [\hat{h}_2^n,y_1^m]=y_1^{n+m+1}+4y_1^{n+m}, &
[\hat{h}_2^n,x_2^m]=2x_2^{n+m+1}+8x_2^{n+m}, \\
 \br\hat{h}_2^n,y_2^m]=-2y_2^{n+m+1}-8y_2^{n+m},&
[\hat{h}_2^n,x_3^m]=x_3^{n+m+1}+4x_3^{n+m}, & [\hat{h}_2^n,y_3^m]=-y_3^{n+m+1}-4y_3^{n+m}, \\
\br x_1^n,x_2^m]=x_3^{n+m}, & [y_1^n,y_2^m]=-y_3^{n+m}, & [x_1^n,y_1^m]=\hat{h}_1^{n+m}, \\
 \br x_1^n,y_3^m]=-y_2^{n+m+1},&
[x_2^n,y_2^m]=\hat{h}_2^{n+m}, & [x_2^n,y_3^m]=y_1^{n+m+1}+4y_1^{n+m}, \\ 
\br x_3^n,y_1^m]=-x_2^{n+m+1}, & [x_3^n,y_2^m]=x_1^{n+m+1}+4x_1^{n+m}, & 
[x_3^n,y_3^m]=\hat{h}_1^{n+m+1}+\hat{h}_2^{n+m+1}+4\hat{h}_1^{n+m}.
\end{array}
\ed}
Elements $x_i^n, y_i^n, \hat{h}_j^n$ with $n\ge 0$ form a basis in $\hat{sl}_{\bbbd_2^A}(3,\bbbc; 0)$, elements with $n<0$ form a basis in $\hat{sl}_{\bbbd_2^A}(3,\bbbc;1)$.

{\em Case A,} $G=\bbbd_3^A$: The action of the reduction group $G=\bbbd_3^A$
can be generated by transformations 
\begin{equation}\label{d3_3}
 g_s: a(\lambda)\to Q_sa(\omega^{-1} \lambda)Q_s^{-1} \, ,\quad g_t: a(\lambda)\to Q_t a(1/\lambda)Q_t ^{-1}\, ,
\end{equation}
where 
\[Q_s=\left( \begin{array}{ccc}
\omega & 0&0\\
0&\omega^2&0\\
0&0&1 \end{array}\right) \, ,\qquad Q_t =\left( \begin{array}{ccc}
0 & 1&0\\
1&0&0\\
0&0&\mp 1 \end{array}\right), \quad \omega=\exp\left(\frac{2\pi i}{3}\right)
\] 
It is easy to check that $g_s^3=g_t^2=(g_s g_t)^2=id$. The signs in $Q_t$ correspond to two inequivalent representations of $\bbbd_3^A$. 

Let us choose the following automorphic function
\begin{equation}\label{primf}
f=\lambda^3 +\lambda^{-3}\,,
\end{equation}
corresponding to the orbits $\bbbd_3 (0)=\{0,\infty\}$ and $\bbbd_3 (\varpi) $  where $\varpi=\exp (\pi i /6)$. The automorphic Lie algebra $sl_{\bbbd_3^A}(3,\bbbc; 0,\varpi)=\bigoplus_{n\in \bbbz}\cA_n$ is quasigraded. A basis in the eight dimensional space $\cA_0$ can be chosen as:
\begin{equation}\label{basis33} \begin{array}{ll}
x_1^0=\langle 2\e_{12}\lambda^{-1}\rangle_{\bbbd_3^A}=\lambda^{-1}\e_{12}+\lambda \e_{21}\, ,&
y_1^0=\langle 2\e_{21}\lambda^{-2}\rangle_{\bbbd_3^A}=\lambda^{-2}\e_{21}+\lambda^2\e_{12}\\
x_2^0=\langle 4\e_{23}\lambda^{-1}\rangle_{\bbbd_3^A}=2\lambda^{-1}\e_{23}\mp 2\lambda \e_{13}\, ,&
y_2^0=\langle 4\e_{32}\lambda^{-2}\rangle_{\bbbd_3^A}=2\lambda^{-2}\e_{32}\mp\lambda^2\e_{31}\\
x_3^0=[x_1^0,x_2^0]=2\lambda^{-2}\e_{13}\pm 2\lambda^{2}\e_{23}\, , &
y_3^0=\langle 4\e_{31}\lambda^{-1}\rangle_{\bbbd_3^A}=2\lambda^{-1}\e_{31}\mp 2\lambda\e_{32}\end{array}
\end{equation}
\begin{equation}\label{basis33h}
 h_1^0=\langle 2(\e_{11}-\e_{22})\lambda^{-3}\rangle_{\bbbd_3^A}=(\lambda^{-3}-\lambda^3)(\e_{11}-\e_{22})\, ,\quad h_2^0=\frac{2}{3}(\e_{11}+\e_{22}-2\e_{33})\, .
\end{equation}

In the basis 
\begin{equation}\label{basis33n}
x_i^n=f^{n}x_i^0\, ,\quad y_i^n= f^{n }y_i^0\, ,\quad  h_j^n= f^{n }h_j^0\, ,\quad i\in\{1,2,3\},\quad j\in\{1,2\},\quad  n\in \bbbz
\end{equation}
of the automorphic Lie algebra $sl_{\bbbd_3^A}(3,\bbbc; 0,\varpi)$ the non-vanishing commutation relations are ($n,m\in\bbbz$)
{\footnotesize\bd \begin{array}{lll}
\left[h_1^n,x_1^m\right]=2x_1^{n+m+1}-4y_1^{n+m}, & \left[h_1^n,y_1^m\right]=-2y_1^{n+m+1}+4x_1^{n+m}, & \left[h_1^n,x_2^m\right]=-x_2^{n+m+1}\mp 2x_3^{n+m},\\ \left[h_1^n,y_2^m\right]=y_2^{n+m+1}\pm 2y_3^{n+m}, & \left[h_1^n,x_3^m\right]=x_3^{n+m+1}\pm 2x_2^{n+m}, & \left[h_1^n,y_3^m\right]=-y_3^{n+m+1}\mp 2y_2^{n+m}, \\
\left[h_2^n,x_2^m\right]=2x_2^{n+m}, & \left[h_2^n,y_2^m\right]=-2y_2^{n+m},&
\left[h_2^n,x_3^m\right]=2x_3^{n+m}, \\
 \left[h_2^n,y_3^m\right]=-2y_3^{n+m}, &
\left[x_1^n,x_2^m\right]=x_3^{n+m}, & \left[x_1^n,x_3^m\right]=x_2^{n+m}, \\
 \left[y_1^n,y_2^m\right]=-y_3^{n+m+1}\mp y_2^{n+m},&
\left[y_1^n,y_3^m\right]=\pm y_3^{n+m}, &
\left[x_1^n,y_1^m\right]=h_1^{n+m}, \\
 \left[x_1^n,y_2^m\right]=-y_3^{n+m},& \left[x_1^n,y_3^m\right]=-y_2^{n+m},&
\left[x_2^n,y_1^m\right]=\pm x_2^{n+m}, \\
\left[x_2^n,y_2^m\right]=3h_2^{n+m+1}-2h_1^{n+m}\mp 4x_1^{n+m}, & \left[x_2^n,y_3^m\right]=\mp 6 h_2^{n+m}+4y_1^{n+m}, &
\left[x_3^n,y_1^m\right]=-x_2^{n+m+1}\mp x_3^{n+m}, \\
\left[x_3^n,y_2^m\right]=4x_1^{n+m+1}-4y_1^{n+m}\mp 6 h_2^{n+m},&
\left[x_3^n,y_3^m\right]=3 h_2^{n+m+1}+2h_1^{n+m}\mp 4x_1^{n+m}.&
\end{array}
\ed}

A subset of (\ref{basis33n}) with $n\ge 0$ form a basis of the subalgebra  $sl_{\bbbd_3^A}(3,\bbbc; 0)$, while elements with $n< 0$ are a basis of the subalgebra $sl_{\bbbd_3^A}(3,\bbbc; \varpi)$, and it follows from the above commutation relations that algebra $sl_{\bbbd_3^A}(3,\bbbc; 0,\varpi)$ is a direct sum of these subalgebras.

\subsubsection{Case B. Inner and outer automorphisms representation.}

Reduction groups $\bbbd_2^B$ and $\bbbd_N^B$ with  $N>2$ yield different automorphic Lie algebras and we consider these sub-cases separately. In the both sub-cases we shall use a primitive automorphic function $f=\lambda^N+\lambda^{-N}$.

{\em Case B,} $G=\bbbd_2^B$: The action of the reduction group $G=\bbbd_2^B$
can be generated by two transformations 
\begin{equation}\label{d2_3out}
 g_s: a(\lambda)\to Q_sa(-\lambda)Q_s^{-1} \, ,\quad g_t: a(\lambda)\to - a^{tr}(1/\lambda)\, ,
\end{equation}
where $Q_s=\mbox{diag}\, (-1,-1,1)$. Indeed, these transformations generates the group $\bbbd_2$, is easy to check that $g_s^2=g_t^2=(g_s g_t)^2=id$.  If one ignores the $\lambda$ transformations (i.e. takes $\pi_2$ natural projection), then the first transformations in (\ref{d2_3out}) is an inner automorphism of algebra $sl(3,\bbbc)$, while the second one is an outer automorphism.

The corresponding automorphic Lie algebra $sl_{\bbbd_2^B}(3,\bbbc; 0,\exp( i\pi/4))$ has a basis of the form (\ref{basis33n}) where 
\beq\begin{array}{lll}
x_1^0=\langle 2\e_{12}\rangle_{\bbbd_2^B}=\e_{12}-\e_{21}\, ,&
y_1^0=\langle 2\e_{21}\l^{-2}\rangle_{\bbbd_2^B}=\lambda^{-2}\e_{21}-\lambda^2\e_{12}\, ,\\
x_2^0=\langle 2\e_{23}\lambda^{-1}\rangle_{\bbbd_2^B}=\lambda^{-1}\e_{23}-\lambda \e_{32}\, ,&
y_2^0=\langle 2\e_{32}\lambda^{-1}\rangle_{\bbbd_2^B}=\lambda^{-1}\e_{32}-\lambda \e_{23}\, ,\\
x_3^0=\langle 2\e_{13}\lambda^{-1}\rangle_{\bbbd_2^B}=\lambda^{-1}\e_{13}-\lambda \e_{31}\, ,&
y_3^0=\langle 2\e_{31}\lambda^{-1}\rangle_{\bbbd_2^B}=\lambda^{-1}\e_{31}-\lambda \e_{13}\, ,
\end{array}
\end{equation}
\[
 h_1^0=\langle 2(\e_{11}-\e_{22})\lambda^{-2}\rangle_{\bbbd_2^B}=(\lambda^{-2}-\lambda^2)(\e_{11}-\e_{22})\, ,\quad  h_2^0=\langle 2(\e_{22}-\e_{33})\lambda^{-2}\rangle_{\bbbd_2^B}=(\lambda^{-2}-\lambda^2)(\e_{22}-\e_{33})\, .\]
The nonvanishing commutation relations of the automorphic Lie algebra $sl_{\bbbd_2^B}(3,\bbbc; 0,\exp \pi i/4)$ are
{\small\bd \begin{array}{lll}
\left[h_1^n,x_1^m\right]=2x_1^{n+m+1}+4y_1^{n+m}, & \left[h_1^n,y_1^m\right]=-2y_1^{n+m+1}-4x_1^{n+m}, & \left[h_1^n,x_2^m\right]=-x_2^{n+m+1}-2y_2^{n+m}, \\
 \left[h_1^n,y_2^m\right]=y_2^{n+m+1}+ 2x_2^{n+m}, & \left[h_1^n,x_3^m\right]=x_3^{n+m+1}+ 2y_3^{n+m}, & \left[h_1^n,y_3^m\right]=-y_3^{n+m+1}- 2x_3^{n+m}, \\
\left[h_2^n,x_1^m\right]=-x_1^{n+m+1}-2y_1^{n+m}, & \left[h_2^n,y_1^m\right]=y_1^{n+m+1}+2x_1^{n+m},&
\left[h_2^n,x_2^m\right]=2x_2^{n+m+1}+4y_2^{n+m}, \\
 \left[h_2^n,y_2^m\right]=-2y_2^{n+m+1}-4x_2^{n+m},&
\left[h_2^n,x_3^m\right]=x_3^{n+m+1}+2y_3^{n+m}, & \left[h_2^n,y_3^m\right]=-y_3^{n+m+1}-2x_3^{n+m}, \\
\left[x_1^n,y_1^m\right]=h_1^{n+m}, & \left[x_1^n,y_2^m\right]=y_3^{n+m},&\left[x_1^n,y_3^m\right]=-y_2^{n+m},\\
\left[x_1^n,x_2^m\right]=x_3^{n+m}, & \left[x_1^n,x_3^m\right]=-x_2^{n+m}, &\left[x_2^n,y_1^m\right]=-y_3^{n+m}, \\
\left[x_2^n,y_2^m\right]=h_2^{n+m}, &\left[x_2^n,y_3^m\right]=y_1^{n+m}, &\left[x_2^n,x_3^m\right]=x_1^{n+m}, \\
\left[x_3^n,y_1^m\right]=-x_2^{n+m+1}-y_2^{n+m}, & \left[x_3^n,y_2^m\right]=x_1^{n+m+1}+y_1^{n+m},&
\left[x_3^n,y_3^m\right]= h_2^{n+m}+h_1^{n+m},\\
\left[y_1^n,y_2^m\right]=-y_3^{n+m+1}- x_3^{n+m},&
\left[y_1^n,y_3^m\right]=-x_2^{n+m}, & \left[y_2^n,y_3^m\right]=x_1^{n+m}.
\end{array}
\ed}

{\em Case B,} $G=\bbbd_3^B$: The action of the reduction group $G=\bbbd_3^B$
can be generated by transformations 
\begin{equation}\label{d3_3out} g_s: a(\lambda)\to Q_sa(\omega^{-1}\lambda)Q_s^{-1} \, ,\quad g_t: a(\lambda)\to - a^{tr}(1/\lambda)\, ,
\end{equation}
where $Q_s=\mbox{diag}\, (\omega ,\omega^2 ,1)$.

A basis of the algebra $sl_{\bbbd_3^B}(3,\bbbc; 0,\exp \pi i/6)$ has of the form (\ref{basis33n}) where 
\begin{equation}\label{basis33b} \begin{array}{ll}
x_1^0=\langle 2\e_{12}\lambda^{-2}\rangle_{\bbbd_3^B}=\lambda^{-2}\e_{12}-\lambda^{2}\e_{21}\, ,&
y_1^0=\langle 2\e_{21}\lambda^{-1}\rangle_{\bbbd_3^B}=\lambda^{-1}\e_{21}-\lambda \e_{12}\, ,\\
x_2^0=\langle 2\e_{23}\lambda^{-2}\rangle_{\bbbd_3^B}=\lambda^{-2}\e_{23}-\lambda^2 \e_{32}\, ,&
y_2^0=\langle 2\e_{32}\lambda^{-1}\rangle_{\bbbd_3^B}=\lambda^{-1}\e_{32}-\lambda \e_{23}\, ,\\
x_3^0=\langle 2\e_{13}\lambda^{-1}\rangle_{\bbbd_3^B}=\lambda^{-1}\e_{13}-\lambda \e_{31}\, ,&
y_3^0=\langle 2\e_{31}\lambda^{-2}\rangle_{\bbbd_3^B}=\lambda^{-2}\e_{31}-\lambda^2 \e_{13}\, ,
\end{array}
\end{equation}
\[
 h_1^0=\langle 2(\e_{11}-\e_{22})\lambda^{-3}\rangle_{\bbbd_3^B}=(\lambda^{-3}-\lambda^3)(\e_{11}-\e_{22})\, ,\quad  h_2^0=\langle 2(\e_{22}-\e_{33})\lambda^{-3}\rangle_{\bbbd_3^B}=(\lambda^{-3}-\lambda^3)(\e_{22}-\e_{33})\, .\]
The non-vanishing commutation relations of the automorphic Lie algebra $sl_{\bbbd_3^B}(3,\bbbc; 0,\exp \pi i/6)$ are 
{\small\bd \begin{array}{lll}
\left[h_1^n,x_1^m\right]=2x_1^{n+m+1}+4y_1^{n+m}, & \left[h_1^n,y_1^m\right]=-2y_1^{n+m+1}-4x_1^{n+m}, & \left[h_1^n,x_2^m\right]=-x_2^{n+m+1}-2y_2^{n+m}, \\
\left[h_1^n,y_2^m\right]=y_2^{n+m+1}+ 2x_2^{n+m}, & \left[h_1^n,x_3^m\right]=x_3^{n+m+1}+ 2y_3^{n+m}, & \left[h_1^n,y_3^m\right]=-y_3^{n+m+1}- 2x_3^{n+m}, \\
\left[h_2^n,x_1^m\right]=-x_1^{n+m+1}-2y_1^{n+m}, & \left[h_2^n,y_1^m\right]=y_1^{n+m+1}+2x_1^{n+m},&
\left[h_2^n,x_2^m\right]=2x_2^{n+m+1}+4y_2^{n+m}, \\
 \left[h_2^n,y_2^m\right]=-2y_2^{n+m+1}-4x_2^{n+m},&
\left[h_2^n,x_3^m\right]=x_3^{n+m+1}+2y_3^{n+m}, & \left[h_2^n,y_3^m\right]=-y_3^{n+m+1}-2x_3^{n+m}, \\
\left[x_1^n,y_1^m\right]=h_1^{n+m}, & \left[x_1^n,y_2^m\right]=-x_3^{n+m},&\left[x_1^n,y_3^m\right]=-y_2^{n+m+1}-x_2^{n+m},\\
\left[x_1^n,x_2^m\right]=x_3^{n+m+1}+y_3^{n+m}, & \left[x_1^n,x_3^m\right]=y_2^{n+m}, &\left[x_2^n,y_1^m\right]=x_3^{n+m}, \\
\left[x_2^n,y_2^m\right]=h_2^{n+m}, &\left[x_2^n,y_3^m\right]=y_1^{n+m+1}+x_1^{n+m}, &\left[x_2^n,x_3^m\right]=-y_1^{n+m}, \\
\left[x_3^n,y_1^m\right]=-x_2^{n+m}, & \left[x_3^n,y_2^m\right]=x_1^{n+m},&
\left[x_3^n,y_3^m\right]= h_2^{n+m}+h_1^{n+m},\\
\left[y_1^n,y_2^m\right]=-y_3^{n+m},&
\left[y_1^n,y_3^m\right]=+y_2^{n+m}, & \left[y_2^n,y_3^m\right]=-y_1^{n+m}.
\end{array}
\ed}

It follows from the commutation relations  that the automorphic Lie algebra $sl_{\bbbd_2^B}(3,\bbbc; 0,\exp \pi i/4)$  (similarly $sl_{\bbbd_3^B}(3,\bbbc; 0,\exp \pi i/6)$) is a direct sum of two subalgebras $sl_{\bbbd_2^B}(3,\bbbc; 0)$ and $sl_{\bbbd_2^B}(3,\bbbc;\exp \pi i/4)$ (correspondingly $sl_{\bbbd_3^B}(3,\bbbc; 0)$ and $sl_{\bbbd_3^B}(3,\bbbc;\exp \pi i/6)$). Basis elements with non-negative upper index form a basis of   $sl_{\bbbd_2^B}(3,\bbbc; 0)$ ($sl_{\bbbd_3^B}(3,\bbbc; 0)$), while elements with negative index are a basis of $sl_{\bbbd_2^B}(3,\bbbc;\exp \pi i/4)$ ($sl_{\bbbd_3^B}(3,\bbbc;\exp \pi i/6)$).
This algebra does not have constant ($\lambda$ independent) elements.

\subsection{Automorphic Lie algebras corresponding to {\em twisted} ($\lambda$-dependent) automorphisms.}\label{semidirect}

In the previous sections we assumed that elements of the group $\aut \cA$ do not depend on the complex parameter $\lambda$. The group $\aut \cA$ is a continuous Lie group and we can admit that some of its elements depend on $\lambda$. In this case, the transformations of a reduction group $G$ can be represented by
pairs $(\sigma, \psi (\lambda))$, where the first element of the pair is a fractional-linear  transformation of the complex plane $\lambda$ while the second entry is a ``$\lambda$--dependent'' automorphism of the Lie algebra $\cA$. To treat this case one needs to generalise the direct product of groups to the \emph{semi-direct product} of groups \cite{hoh}, \cite{coh}:
\begin{df} Let $G_1\, ,\, G_2$ be two groups and $\phi$ be a homomorphism of $G_1$ into the group of automorphisms  of $G_2$, denoted by $\aut G_2$. Then $G_1\times G_2$ with the product defined by 
\[ (x,y)\cdot (x_1,y_1)=(x\cdot x_1,y\cdot \phi(x)y_1) \] 
is a group called the {\em semi-direct product} and denoted by $G_1\times _\phi G_2$.
\end{df}
When the homomorphism $\phi:G_1\to \aut G_2$ is such that  $\phi (x)$ is the identity
(i.e. $\phi (x)y=y,\, \forall x\in G_1,\forall  y\in G_2$), then we obtain the direct product. 
It is easy to verify that $H_1=\{id\}\times_\phi G_2=\{(id, x)\, |\, x\in G_2\}$ is a normal subgroup of $G_1\times _\phi G_2$, while the subgroup $H_2=G_1\times_\phi \{id\}=\{(x,id)\, |\, x\in G_1\}$ is not necessarily normal. Therefore Theorem \ref{th1} is not valid for the semi-direct product.

The composition rule for ``$\lambda$ dependent'' elements of a reduction group is
similar to the rule for a semi-direct product of the groups $\aut \cR(\Gamma)$ and $\aut \cA$. Indeed, it is easy to show that 
\[ (\sigma, \psi(\lambda))=(\sigma_2,\psi_2(\lambda))\cdot(\sigma_1,\psi_1(\lambda))=
(\sigma_2\cdot\sigma_1,\psi_2(\lambda)\cdot \sigma_2(\psi_1(\lambda)))\, .\]
In this case the homomorphism $\phi:\aut\cR(\Gamma)\to\aut(\aut \cA)$ is the corresponding fractional linear transformation of parameter $\lambda$. 

Let us consider a nontrivial example of a reduction group $\bbbd_3^\lambda\cong\bbbd_3$ with $\lambda$ dependent automorphisms of $sl(3,\bbbc)$ and the corresponding infinite dimensional automorphic Lie algebra. Let  $\bbbd_3^\lambda$ be a group of transformations generated by
\[ g_s:a(\lambda)\to Qa(\omega^{-1} \lambda)Q^{-1}\, ,\quad g_t:a(\lambda)\to -T(\lambda)a^{tr}(\lambda^{-1})T^{-1}(\lambda)\, ,\quad a(\lambda)\in sl(3,\bbbc)\]
where
\[ Q=\left( \begin{array}{ccc}
\omega&0&0\\
0&\omega^2&0\\
0&0&1\end{array}\right) \, ,\quad  T(\lambda)=\frac{\lambda^3}{1-\lambda^6}\left( \begin{array}{ccc}
1&\lambda^2&\lambda^{-2}\\
\lambda^{-2}&1&\lambda^{2}\\
\lambda^{2}&\lambda^{-2}&1\end{array}\right)\,
,\quad  T^{-1}(\lambda)=\left( \begin{array}{ccc}
0&\lambda^{-1}&-\lambda\\
-\lambda&0&\lambda^{-1}\\
\lambda^{-1}&-\lambda&0\end{array}\right) .\]
It is easy to check that $g_s^3=id$. Also one can check that $g_t^2=id$. Indeed, since $T(\lambda)(T^{-1}(\lambda^{-1}))^{tr}=-I$ we have 
{\small\[ g_t\cdot g_t : a(\lambda)\to -T(\lambda)\left(-T(\lambda^{-1})a^{tr}(\lambda)T^{-1}(\lambda^{-1})\right)^{tr}T^{-1}(\lambda)=
T(\lambda)(T^{-1}(\lambda^{-1}))^{tr}a(\lambda)(T(\lambda^{-1}))^{tr}T^{-1}(\lambda)=a(\lambda). \]}
Similarly, one can check that $g_s\cdot g_t\cdot g_s\cdot g_t=id$. Thus, the group $\bbbd_3^\lambda\cong\bbbd_3$.

Let us describe the space of $\bbbd_3^\lambda$ invariant $3\times 3$ matrices with rational entries in $\lambda$ and with simple, double and third order poles at points $\{0,\infty\}$. Matrix $a(\lambda)$ is $\bbbd_3^\lambda$ invariant if and only if 
\begin{equation}\label{invcond}
a(\lambda)= Qa(\omega^{-1}\lambda)Q^{-1}\, ,\qquad a(\lambda)= -T(\lambda)a^{tr}(\lambda^{-1})T^{-1}(\lambda)\, .
\end{equation}

\begin{prop} The zero matrix is the only constant and $\bbbd_3^\lambda$ invariant. If a matrix is rational in $\lambda$ with poles at $\{0,\infty\}$ and $\bbbd_3^\lambda$ invariant, then it can be uniquely represented as a linear combination of:
\begin{enumerate}
\item in the case of simple poles 
\begin{equation}\label{xgens}
x_1(\lambda)=\e_{12}\lambda^{-1}-\e_{13}\lambda\, ,\quad x_2(\lambda)=\e_{23}\lambda^{-1}-\e_{21}\lambda\, ,\quad 
x_3(\lambda)=\e_{31}\lambda^{-1}-\e_{32}\lambda\, ;
\end{equation}
\item in the case of double poles 
\begin{equation}\label{ygens}
y_1(\lambda)=[x_2(\lambda),x_3(\lambda)]\, ,\quad y_2(\lambda)=[x_3(\lambda),x_1(\lambda)]\, ,\quad 
y_3(\lambda)=[x_1(\lambda),x_2(\lambda)]\, ,
\end{equation}
and $x_i(\lambda)$ listed in (\ref{xgens});
\item in the case of third order poles 
\begin{equation}\label{zgens}
z_1(\lambda)=[x_1(\lambda),y_1(\lambda)]\, ,\quad z_2(\lambda)=[x_2(\lambda),y_2(\lambda)]\, ,\quad 
z_3(\lambda)=(\lambda^3-\lambda^{-3})I\, ,
\end{equation}
and $x_i(\lambda)\, ,y_i(\lambda)$ listed in (\ref{xgens}), (\ref{ygens}).
\end{enumerate}
\end{prop}
 
\pf If matrix $a$ is constant, then it follows from the first condition (\ref{invcond})
that $a$ is diagonal. The second condition means that the constant, diagonal matrix $a$ anti-commutes with $T(\lambda)$, which is impossible if $a\ne 0$. If the matrix $a(\lambda)$ has simple poles 
at $\{0,\infty\}$, it can be represented as $a(\lambda)=a_0+\lambda a_+ +\lambda^{-1}a_-$, where $a_0,\,a_\pm$ are constant complex matrices. From the first condition (\ref{invcond}) it follows that
\[ a(\lambda) =\left( \begin{array}{ccc}
a_{11}&\lambda^{-1}a_{12}&\lambda a_{13}\\
\lambda a_{21}&a_{22}&\lambda^{-1} a_{23}\\
\lambda^{-1} a_{31}&\lambda a_{32}& a_{33}
\end{array}\right) \, ,\qquad a_{ij}\in \bbbc .\]
The second condition (\ref{invcond}) can be rewritten as $a(\lambda)T(\lambda)+T(\lambda)a^{tr}(\lambda^{-1})=0$
and it is equivalent to a system of linear, homogeneous equations for constant entries $a_{ij}$.
This system has three nontrivial solutions which can be written in the form (\ref{xgens}).
In the case of second order poles we represent $a(\lambda)$ as $a_0+\lambda a_+ +\lambda^{-1}a_-+\lambda^2 b_+ +\lambda^{-2}b_-$. Conditions (\ref{invcond}) yield to a system of linear equations for the constant matrices $a_0,\,a_\pm,\,b_\pm$, whose general solution can be written in the form $a(\lambda)=\sum_{i=1}^3\alpha_i y_i(\lambda)+\beta_i x_i(\lambda)\, ,\ \alpha_i,\beta_i\in\bbbc$. The case of third order poles can be  treated similarly. $\square $ 

\begin{prop}
\begin{enumerate}
\item The set 
\begin{equation}\label{lbasis} 
\{x_i^n=x_i(\lambda) f^n \, , y_i^n=y_i(\lambda) f^n \, , h_j^n=z_j(\lambda) f^n \, |\,  
i\in\{1,2,3\},\ j\in\{1,2\},\ n\in\bbbz\, ,f=\lambda^3+\lambda^{-3}\}, \end{equation}
is a basis of the automorphic Lie algebra  $sl_{\bbbd_3^\lambda}(3,\bbbc;0,\exp (\pi i/6))$.
\item $sl_{\bbbd_3^\lambda}(3,\bbbc;0,\exp (\pi i/6))$ is a direct sum of two subalgebras 
$sl_{\bbbd_3^\lambda}(3,\bbbc;0)$ and $sl_{\bbbd_3^\lambda}(3,\bbbc;\exp (\pi i/6))$.
\item The subsets $\{x_i^n,y_i^n,z_j^n\, |\, n\ge 0\}$ and $\{x_i^n,y_i^n,z_j^n\, |\, n<0\}$
of the set (\ref{lbasis}) are bases of subalgebras $sl_{\bbbd_3^\lambda}(3,\bbbc;0)$ and $sl_{\bbbd_3^\lambda}(3,\bbbc;\exp (\pi i/6))$, respectively.
\item $sl_{\bbbd_3^\lambda}(3,\bbbc;0)$ is generated by $x_1(\lambda),x_2(\lambda)$ and $x_3(\lambda)$.
\end{enumerate}
\end{prop}

\pf The proof of the first statement of the Proposition is similar to the proofs of Proposition \ref{propbasis} and Proposition \ref{negbasis}. The proof of the rest follows from the commutation relations for the basis elements of the algebra
\begin{equation}\label{cr1}
[x_i^n,x_j^m]=\epsilon_{ijk}y_k^{n+m},\,\,
[y_i^n,y_j^m]=-\epsilon_{ijk}(x_k^{n+m+1}-y_i^{n+m}-y_j^{n+m}), \,\,
[x_i^n,y_j^m]=-2\epsilon_{ijk}x_i^{n+m},\quad  i\ne j,
\end{equation}
\begin{equation}\label{cr2}
[x_1^n,y_1^m]=h_1^{n+m},\quad  [x_2^n,y_2^m]=h_2^{n+m},\quad  [x_3^n,y_3^m]=-h_1^{n+m}-h_2^{n+m}, \end{equation}
\begin{equation}\label{cr3}
[h_2^n,h_1^m]=\sum_{k=1}^3 (x_k^{n+m+1}-2y_k^{n+m}),\quad [h_i^n,x_i^m]=2x_i^{n+m+1},\quad i=1,2\, ,
\end{equation}
\begin{equation}\label{cr4}
\ba{l}
 \left[h_1^n,y_1^m\right]=-h_1^{n+m}-2h_2^{n+m}+2(x_2^{n+m}+x_3^{n+m}-y_1^{n+m+1})\, ,\\
 \left[h_2^n,y_2^m\right]=2h_1^{n+m}+h_2^{n+m}+2(x_1^{n+m}+x_3^{n+m}-y_2^{n+m+1})\, ,
\ea
\end{equation}
\begin{equation}\label{cr5}
[h_i^n,x_j^m]=-x_j^{n+m+1}+y_i^{n+m}-|\epsilon_{ijk}|y_k^{n+m},
\quad [h_i^n,y_j^m]=-y_j^{n+m+1}-2x_i^{n+m}-\epsilon_{ijk}h_i^{n+m},\quad  i\ne j.
 \end{equation}

Elements with non-negative upper index form a closed $\bbbd_3^\lambda$--automorphic subalgebra and they have poles at points $\{0,\infty\}$.  This subalgebra is generated by $x_i^0=x_i(\lambda)$. Indeed, $y_i^0$ can be found from (\ref{cr1}), $h_i^0$ from (\ref{cr2}), $x_i^1,y_i^1$ from (\ref{cr4}), etc.

Elements with negative upper index 
also form a closed subalgebra, they have poles at the points of the orbit $\{\exp\left(\frac{(2n+1)\pi i}{6}\right)\, |\, n=1,...,6\}$ and are regular elsewhere. $\square$

Algebra $sl_{\bbbd_3^\lambda}(3,\bbbc;0)$  has been discovered in \cite{msy1}, but
its automorphic nature and the reduction group was not known
until now. It is not difficult to show that it does not exist any $\lambda$
independent reduction group which corresponds to $sl_{\bbbd_3^\lambda}(3,\bbbc;0)$.

\appendix

\section{Appendix. Finite groups of fractional-linear transformations, their orbits and primitive automorphic functions}

\subsubsection*{The group $\bbbz_N$}

The group $\bbbz_N$ can be represented by the following transformations
\begin{equation}\label{ZN}
\sigma_n(\lambda)=\Omega^n \lambda\, ,\qquad \Omega=\exp\left(\frac{2\pi i}{N}\right)\, ,\qquad n=0,1,\ldots,N-1\, .
\end{equation}
It has two degenerated orbits $\bbbz_N(0)=\{0\}$, $\bbbz_N(\infty)=\{\infty\}$ corresponding to  
two fixed points of order $N$ and a generic orbit 
$ \bbbz_N(\gamma)=\{\gamma,\Omega \gamma,\Omega^2\gamma,\ldots ,\Omega^{N-1}\gamma\}\, ,\ \gamma\not\in\{0,\infty\}$. 
A primitive automorphic 
function, corresponding to the orbits $\bbbz_N(0)$, $\bbbz_N(\infty)$ is
\[ f_{\bbbz_N}(\lambda,\infty,0)=\lambda^N\, .\]  

It follows from (\ref{gam21})--(\ref{gam34}) that for $\gamma_1\ne\infty$ and $\gamma_2\not\in \bbbz_N(\gamma_1)$
\[ f_{\bbbz_N}(\lambda,\gamma_1,\gamma_2)=\frac{\lambda^N-\gamma_2^N}{\lambda^N-\gamma_1^N}\, .\]  

\subsubsection*{The dihedral group $\bbbd_N$}

The group $\bbbd_N$ has order $2N$ and can be generated by the 
following transformations
\begin{equation}\label{DN}
\sigma_s (\lambda)=\Omega \lambda\, ,\qquad \sigma_t (\lambda)=\frac{1}{ \lambda} \, 
,\qquad\Omega=\exp\left(\frac{2\pi i}{N}\right)\, .
\end{equation}
Transformations $\sigma_s,\sigma _b$ satisfy the relations $\sigma_s^N=\sigma_t^2=(\sigma_s\sigma_t)^2=id$
and
\[ \bbbd_N=\{\sigma_s^n,\sigma_s^n\sigma_t\, |\, \, n=0,\, \ldots \, ,N-1\}\, .\]

For $N\ge 3$ the group $\bbbd _N$ is non-commutative, the case $N=2$ is special,
in this case $\bbbd _2\cong \bbbz_2 \times \bbbz_2$ and it is commutative. F.~Klein calls it the quadratic group (some authors call $\bbbd _2$ the Klein group).

The group $\bbbd _N$ has three degenerated orbits and one generic orbit. The structure of the orbits is different for odd and even $N$.
For odd $N$ we have:
\begin{eqnarray}\label{DN0}
\bbbd_N(0)&=&\{0,\infty\}\, , \\ \label{DN1}
\bbbd_N(1)&=&\{\ 1,\Omega,\, \ldots\, ,\Omega^{N-1}\}\, ,\\ \label{DN-1}
\bbbd_N(-1)&=&\{\ -1,-\Omega,\, \ldots\, ,-\Omega^{N-1}\}\, ,\\ \label{DNgamma}
\bbbd_N(\gamma)&=&\{\ \gamma\, ,\Omega\gamma\, , \ldots\, ,\Omega^{N-1}\gamma\, ,
\gamma^{-1}\, ,\Omega\gamma^{-1}\, , \ldots\, ,\Omega^{N-1}\gamma^{-1}\}\, .
\end{eqnarray}
For even $N$ orbits $\bbbd_N(1)$ and $\bbbd_N(-1)$ coincide and instead of (\ref{DN-1}) we have the orbit
\begin{equation} 
\bbbd_N(i)=\{\ i,i\Omega,\, \ldots\, ,i\Omega^{N-1}\}\, . \label{DNi} 
\end{equation}
The orbit (\ref{DN0}) consists of fixed points of order $N$. Orbits (\ref{DN1}), (\ref{DN-1}), (\ref{DNi})
consist of fixed points of order $2$ (they correspond to the vertices of the dihedron or to the middles
of the edges, i.e. vertices of the dual dihedron).  
A primitive automorphic 
function, corresponding to the orbits $\bbbd_N(0)$, $\bbbd_N(1)$ is
\[ f_{\bbbd_N}(\lambda,0,1)=\lambda^N+\lambda^{-N}-2\, .\]  

\subsubsection*{The tetrahedral group $\bbbt$}

The group of a tetrahedron $\bbbt$ has order $12$ and can be generated by two transformations
\[ \sigma_s(\lambda)=-\lambda\, ,\quad \sigma_t(\lambda)=\frac{\lambda+i}{\lambda-i} \, .\]
It is easy to check that $\sigma_s^2=\sigma_t^3=(\sigma_s\sigma_t)^3=id$
and
\[ \bbbt=\{\sigma_t^n,\sigma_t^n\sigma_s\sigma_t^m\, |\, \, n,m=0,1,2\}\, .\]
The group $\bbbt$ has four distinct orbits.
The orbit corresponding to a generic point $\gamma$ is a set of 12 points 
\[ \bbbt(\gamma)=\{\pm\gamma,\pm\gamma^{-1},\pm i \frac{\gamma+1}{\gamma-1},\pm i \frac{\gamma-1}{\gamma+1}
,\pm \frac{\gamma+i}{\gamma-i},\pm \frac{\gamma-i}{\gamma+i}\}\, ,\]
Transformation $\sigma _a$ has two fixed points of order two, namely $\{0,\infty\}$, the corresponding orbit
consists of six points, which correspond to middle of the edges of the tetrahedron
\begin{equation}\label{T2}
\bbbt(0)=\{ 0,\infty,\pm 1,\pm i\}\, .
\end{equation}
There are two orbits with fixed points of order 3. They correspond to the vertices of the tetrahedron and the dual
tetrahedron. Fixed points of the transformation $\sigma_t$ can be used as seeds for these orbits.
It follows from $\gamma=\sigma_t(\gamma)$ that the fixed points are $\gamma_1=(1+i)/(1+\sqrt{3})=\omega+i \bar{\omega}\, ,\ \gamma_2=(1+i)/(1-\sqrt{3})=i\omega+\bar{\omega}$, where $\omega=\exp(2\pi i/3)$ and 
therefore we have two orbits:
\begin{equation}
 \bbbt (\gamma_1 )=\left\{ \pm (\omega+i \bar{\omega}),\pm (\omega-i \bar{\omega})\right\}\, ,\qquad
\bbbt (\gamma_2 )=\left\{ \pm i(\omega+i \bar{\omega}),\pm i(\omega-i \bar{\omega})\right\}\, .
\end{equation}
Points of the orbits $\bbbt (\gamma_1 )$ and $\bbbt (\gamma_2 )$ are roots of the equations $\lambda^4+2(\omega+\bar{\omega})\lambda^2+1=0$ and $\lambda^4-2(\omega+\bar{\omega})\lambda^2+1=0$,
respectively. A primitive automorphic 
function, corresponding to orbits $\bbbt(\gamma_1)$, $\bbbt(\gamma_2)$ is
\[ f_{\bbbt}(\lambda,\gamma_1,\gamma_2)=\left( \frac{\lambda^4+2 (\omega+\bar{\omega})\lambda^2+1}{\lambda^4-2(\omega+\bar{\omega})\lambda^2+1}\right)^3\, .\]  
It follows from (\ref{gam13}) that 
\[ f_{\bbbt}(\lambda,\gamma_1,0)=f_{\bbbt}(\lambda,\gamma_1,\gamma_2)-1=
12(\omega+\bar{\omega}) \frac{\lambda^2(\lambda^4-1)^2}{(\lambda^4-2(\omega+\bar{\omega})\lambda^2+1)^3}\, .\]  

\subsubsection*{The octahedral group $\bbbo$}

The group of an octahedron $\bbbo$ has order $24$ and can be generated by two transformations
\begin{equation}\label{octgen}
 \sigma_s(\lambda)=i\lambda\, ,\quad \sigma_t(\lambda)=\frac{\lambda+1}{\lambda-1} \, .
\end{equation}
It is easy to check that $\sigma_s^4=\sigma_t^2=(\sigma_s \sigma_t)^3=id $ and
\[ \bbbo=\{\sigma_s^n,\sigma_s^n\sigma_t \sigma_s^m,\sigma_s^n\sigma_t 
\sigma_s^2 \sigma_t\, |\, \, n,m=0,1,2,3\}\, .\]

The group $\bbbo$ has also four distinct orbits corresponding to 
\begin{enumerate}
\item[i] the vertices of the octahedron (a fixed point of order $4$ of the transformation 
$\sigma_s$ belongs to this orbit), therefore 
\[ \bbbo(0)=\bbbt(0)\, ; \]
\item[ii] the centres of the triangular faces (i.e. vertices of the cube - the dual to the octahedron).
the point $\gamma_1$, a fixed point of $\sigma_t$, belongs to this orbit, therefore
\[ \bbbo(\gamma_1)=\bbbt(\gamma_1)\bigcup \bbbt(\gamma_2)\, ;\, \]
\item[iii] the middles of the edges of the octahedron 
\[ \bbbo(\delta)=
\{ \pm \delta,\
  \pm \bar{\delta},\ 
  i^n (1+\delta+\bar{\delta}),\
 i^n (1-\delta-\bar{\delta})\, |\, 
  n=0,1,2,3\} \]
where $\delta=\exp(\pi i/4)$  is one of the points on the middle of an edge of the octahedron, for example 
a fixed point of the transformation $\lambda\to i/\lambda$, which 
belongs to the group generated by $\sigma_s, \sigma_t$ (\ref{octgen})); 
\item[iv] the orbit, corresponding to a generic point $\gamma$ (i.e. $\gamma$ does not belong to the above listed orbits)
is a set of 24 points 
\[ \bbbo(\gamma)=\{i^k\gamma,i^k\gamma^{-1},i^k \frac{\gamma+1}{\gamma-1},i^k \frac{\gamma-1}{\gamma+1}
,i^k \frac{\gamma+i}{\gamma-i},i^k \frac{\gamma-i}{\gamma+i}\}\, ,\qquad k\in\{0,1,2,3\}\, .\]
\end{enumerate}

A primitive automorphic 
function, corresponding to orbits $\bbbo(0), \bbbo(\gamma_1)$ is
\[ f_{\bbbo}(\lambda,0,\gamma_1)= \frac{(\lambda^4-2(\omega+\bar{\omega})\lambda^2+1)^3(\lambda^4+2(\omega+\bar{\omega})\lambda^2+1)^3}{\lambda^4(\lambda^4-1)^4}=
\frac{(\lambda^8+14 \lambda^4 +1)^3}{\lambda^4(\lambda^4-1)^4}
\, .\]  

\subsubsection*{The icosahedral group $\bbbi$}

The group of the icosahedron $\bbbi$ has order $60$ and can be generated by two transformations
\begin{equation}\label{icgen}
 \sigma_s(\lambda)=\varepsilon \lambda\, ,\quad \sigma_t(\lambda)=\frac{(\varepsilon^2+\varepsilon^3)\lambda+1}{\lambda-\varepsilon^2-\varepsilon^3} \, ,\quad \varepsilon=\exp\left(\frac{2 \pi i}{5}\right).
\end{equation}

Its generators satisfy the relations $\sigma_s^5=\sigma_t^2=(\sigma_s \sigma_t)^3=id $ and
\[ \bbbi=\{ \sigma_s^n,\ \sigma_s^n \sigma_t \sigma_s^m,\ 
\sigma_s^n \sigma_t \sigma_s^2 \sigma_t \sigma_s^m,\
\sigma_s^n \sigma_t \sigma_s^2 \sigma_t \sigma_s^3 \sigma_t \, |\, \, n,m=0,1,2,3,4\}.\]

The group $\bbbi$ has also four distinct orbits corresponding to 
\begin{enumerate}
\item[i] the vertices of the icosahedron (fixed points of order $5$ of the transformation 
$\sigma_s$ belong to this orbit)
\[ \bbbi (0)=\{ 0,\infty, \varepsilon^{k+1}+\varepsilon^{k-1}, \varepsilon^{k+2}+\varepsilon^{k-2}\, |\, k=0,1,2,3,4\} \, .\]
The finite points of this orbit are all solutions of the equation
$ \lambda (\lambda^{10}+11 \lambda^5-1)=0$.
\item[ii] the centres of the triangular faces (i.e. vertices of dodecahedron - the dual to the icosahedron). The transformation 
\[ \sigma_s^2 \sigma_t\sigma_s^2(\lambda)=\frac{(1+\bar{\varepsilon})\lambda+1}{\lambda-1-\varepsilon}\]
 is of order $3$ and it has fixed points 
\[ \gamma_1=\frac{3+\sqrt{5}+\sqrt{6(5+\sqrt{5})}}{4}=1-\omega\varepsilon-\bar{\omega}\bar{\varepsilon}\, ,\quad 
 \gamma_2=\frac{3+\sqrt{5}-\sqrt{6(5+\sqrt{5})}}{4}=1-\bar{\omega}\varepsilon-\omega\bar{\varepsilon}\,  \]
(here $\omega=\exp(2\pi i/3)$). 
The corresponding orbit $\bbbi (\gamma_1)$ consists of $20$ points, these points are
solutions of the equation \cite{klein}
\[ \lambda^{20}-228\lambda^{15}+494 \lambda^{10}+228\lambda^5+1=0 \, .\]

\item[iii] the middles of the edges of the icosahedron correspond to the 
orbit $\bbbi (i)$. The point $i$ is a fixed point of transformation 
$\sigma_s^2 \sigma_t\sigma_s^3\sigma_t\sigma_s^2 \sigma_t(\lambda)=-1/\lambda$. Points of this 
orbit are solutions of the equation 
\[ \lambda^{30}+522\lambda^{25}-10005\lambda^{20}-10005\lambda^{10}-522\lambda^5+1=0\, .\]

\item[iv] the orbit, corresponding to a generic point $\gamma$ 
(i.e. $\gamma$ does not belong to the above listed orbits) is a set of 
$60$ points 
\[ \bbbi(\gamma)=\{\varepsilon ^n\gamma,\-\frac{\varepsilon^n}{\gamma},\
\varepsilon ^n \frac{(\varepsilon ^3+\varepsilon ^2)\varepsilon ^m \gamma+1}
{\varepsilon ^m \gamma- \varepsilon ^3-\varepsilon ^2},\ 
-\varepsilon ^n \frac{\varepsilon ^m \gamma- \varepsilon ^3-\varepsilon ^2}
{(\varepsilon ^3+\varepsilon ^2)\varepsilon ^m \gamma+1}\, |\, \, n,m=0,1,2,3,4 \}\, .
\]
\end{enumerate}

A primitive automorphic 
function, corresponding to orbits $\bbbi(0)$, $\bbbi(\gamma_1)$ is
\[ f_{\bbbi}(\lambda,0,\gamma_1)= \frac{(\lambda^{20}-228\lambda^{15}+494 \lambda^{10}+228\lambda^5+1)^3}
{\lambda^5 (\lambda^{10}+11 \lambda^5-1)^5}
\, .\]  

It is easy to check that 
\[ f_\bbbi (\lambda,0,i)=f_\bbbi(\lambda,0,\gamma_1)-f_\bbbi(i,0,\gamma_1)=\frac{(\lambda^{30}+522\lambda^{25}-10005\lambda^{20}-10005\lambda^{10}-522\lambda^5+1)^2}
{\lambda^5 (\lambda^{10}+11 \lambda^5-1)^5}\, .\]

\section*{Acknowledgement}
We would like to thank W Crawley-Boevey, Y Kodama, J Schr\"{o}er, T Skrypnyk and V V Sokolov for interesting discussions and useful comments. The initial stage of the work of 
S L was supported by the University of 
Leeds \emph{William Wright Smith} scholarship and successively by a grant of
the Swedish foundation \emph{Blanceflor 
Boncompagni-Ludovisi, 
n\'ee Bildt}, for which S L is most grateful. The work of A M was 
partially supported by RFBR grant 02-01-00431. 



\end{document}